\documentclass[12pt, aps, prd, tightenlines, nofootinbib, longbibliography, superscriptaddress, notitlepage]{revtex4-1}
\usepackage{xspace}
\usepackage{amsfonts, amsmath, amsthm, amssymb, mathrsfs}
\usepackage[dvips]{graphicx}
\usepackage[left=2cm, right=2cm, top=2.5cm, bottom=2cm]{geometry}
\usepackage[pdfencoding=auto,psdextra]{hyperref}
\usepackage{enumerate}
\usepackage{bbm}
\usepackage{xcolor}
\usepackage{soul}
\usepackage{comment}
\usepackage[splitrule]{footmisc}

\begin{document}

\title{Loop effective model for Schwarzschild black hole:\\ a modified $\bar{\mu}$ dynamics}

\author{Mehdi Assanioussi}
\email[]{mehdi.assanioussi@fuw.edu.pl}
\affiliation{Faculty of Physics, University of Warsaw, Pasteura 5, 02-093 Warsaw, Poland.}
\affiliation{II. Institute for Theoretical Physics, University of Hamburg,\\ Luruper Chaussee 149, 22761 Hamburg, Germany.}

\author{Lisa Mickel}
\email[]{lmickel1@sheffield.ac.uk}
\affiliation{II. Institute for Theoretical Physics, University of Hamburg,\\ Luruper Chaussee 149, 22761 Hamburg, Germany.}
\affiliation{School of Mathematics and Statistics, University of Sheffield, Hicks Building,\\ Hounsfield Road, Sheffield S3 7RH, United Kingdom.}

\count\footins = 1000

\begin{abstract}

In this article, we introduce a new effective model for the Kantowski-Sachs spacetime in the context of loop quantum gravity, and we use it to evaluate departures from general relativity in the case of Schwarzschild black hole interior. The model is based on an effective Hamiltonian constructed via the regularized Thiemann identities in the $\bar \mu$-scheme. We show that, in contrast with the $\mu_o$-scheme studied in \cite{mu0}, the classical limit imposes certain alterations of Thiemann identities as well as restrictions on the choice of regulators. Once we define the Hamiltonian, we derive the equations of motion for the relevant variables and proceed with the solving using numerical methods, focusing on a specific choice of $\bar \mu$. We establish that for a Schwarzschild black hole interior, the effective dynamics leads to a resolution of the classical singularity and the emergence of an anti-trapped region bounded by a second Killing horizon. We then perform a comparison of the dynamical trajectories and their properties obtained in the new model and some models present in the literature. We finally conclude with few comments on other choices of the regulators and their consequences.

\end{abstract}

\maketitle

\section{Introduction}
General relativity, despite being an extremely successful theory of gravity, has some shortcomings: singularities occur in the theory, signaling its breakdown and invalidity in certain scenarios. Examples of such singularities are the big bang singularity and the singularity at the center of black holes. 
There has been an ongoing effort to establish a theory of quantum gravity which, among other things, should resolve the issue of singularities. 
There exist various approaches to define a theory of quantum gravity, none of which are currently complete from a theoretical viewpoint, nor are able to make verifiable predictions. Loop quantum gravity (LQG) is one of these approaches \cite{lqgextra,thiemannBook}, and it consists of a canonical quantization of general relativity in its Hamiltonian formulation in terms of the Ashtekar-Barbero variables \cite{ashtekarVariables,barbero1}. 

In LQG, the dynamical equations are often too involved to be solved explicitly, making the identification of physically relevant subsystems and the calculations therein so far impossible. 
It is therefore judicious to test many concepts and techniques of LQG in symmetry reduced systems, where this issue can be overcome.   
The application of LQG-inspired techniques to the cosmological setting, from which the field of Loop Quantum Cosmology (LQC) emerged, has produced promising results, among which is the resolution of the big bang singularity \cite{LQC_review, LQC_maths}. 
It has been shown that such modifications of quantum origin can be captured in the so-called effective models, which are classical models with a modified dynamics that reproduce the semi-classical limit of the quantum theory\cite{LQC1robustness, LQC1improvedDyn, deSitter1, deSitter2, LQC1uniqueMubar, LQCmu01, LQCmu02, taverasLQC, bojowaldLQC, singhVanderslootLQC}. 
The efforts were also extended to the context of black holes, and several LQG-inspired models for the black hole have been introduced, often concluding that the singularity at the center of the black hole can indeed be resolved by applying LQG techniques \cite{Qbh_Ashtekar, Qbh_Modesto1, Qbh_Modesto2, Qbh_Pullin1, Qbh_Pullin2, Qbh_Pullin3, Qbh_Pullin4, Gambini:2020nsf}.
As a complete quantum description of the black hole interior has not yet been achieved, black hole effective models have attracted a lot of interest. 

In the LQG context, effective models for a given spacetime aim at incorporating certain quantum effects, which may arise within a loop quantum model, into the classical framework. They are defined on the classical phase space with modified classical dynamics, different from the one of general relativity. On the one hand, when a quantum theory is available, it often remains that the solutions to the quantum dynamics are not analytically accessible and the semi-classical analysis is often realized only via numerical calculations. Therefore, when present, effective models are very attractive due to their comparative simplicity which allows to probe the whole phase space with relatively little computational effort. 
On the other hand, for systems for which no quantum theory has been fully established, effective models provide a useful tool to study potential quantum effects that may emerge in the quantum theory, and allow to probe the impact of certain ambiguities in model construction on the resulting phenomenology. 

There exist several approaches to construct an effective model for a spherically symmetric spacetime \cite{CS,AOS,BMM, mu0, Bojowald:2018xxu, BenAchour:2018khr, BenAchour:2017jof, Alesci:2018loi, Alesci:2019pbs, Alesci:2020zfi, Kelly:2020uwj, Geiller:2020xze, Blanchette:2020kkk}. 
Although many effective models for the Schwarzschild black hole interior lead to a singularity resolution, the details of the resulting spacetime and its dynamics differ from one model to another. For instance, the effective models based on the straightforward compactification of the connection variables, via point holonomies, predict the appearance of a white hole like horizon in the interior region \cite{CS,AOS,BMM, mu0}. Other effective models, which are based on a semi-classical derivation of the effective Hamiltonian from a quantum theory, do not necessarily lead to that same conclusion. Namely, in the context of quantum reduced loop gravity (QRLG), the second horizon is not present and one obtains an asymptotic de Sitter spacetime instead \cite{ Alesci:2019pbs, Alesci:2020zfi}. 
A similar semi-classical approach in the context of LQG on the other hand, seems to still generate a second horizon \cite{mu0}.
These qualitative differences in the predictions are known to originate from the ambiguities and various choices present in the construction of different effective models, among which are the choice of phase space variables, the choice of the regularization procedures and parameters, and the choice of semi-classical states in quantum theories. However, the question of what exactly are the elements in the construction responsible for such contrast in the phenomenology remains an open one.

The present article presents the construction and analysis of a new loop effective model for the Schwarzschild black hole interior. The construction of the effective Hamiltonian relies on the combination of Thiemann identities for the regularization of the classical Hamiltonian and the incorporation of phase space dependent regularization parameters, a.k.a.\ the $\bar \mu$-scheme. The model represents an extension of the previously studied $\mu_o$-scheme model in \cite{mu0}. The article is organized as follows: after an introduction to the classical treatment of the Schwarzschild black hole interior in the Hamiltonian formulation of general relativity in sec.\,\ref{sec:pre}, 
we present the construction of the new model in sec.\,\ref{sec:newM_mu}, where we also discuss various choices of regulators and their compatibility with the regularization procedure.
In sec.\,\ref{sec:numAnalysis}, we proceed with a numerical analysis of the model to establish the key phenomenological features, then we make a comparison with some of the models in the literature. We conclude with some comments and remarks on the general construction and possible avenues.
Throughout the article we use the Planck units $c=G=\hbar = 1$ and set $\kappa := 8 \pi G$.

\section{Framework of effective loop black hole interior models}\label{sec:pre}

In this section we introduce the classical description of the Schwarzschild black hole interior as a Kantowski-Sachs spacetime in the Hamiltonian formulation, then we discuss the main aspects of black hole effective models which have been studied in the literature, and motivate the model we construct in this article.

\subsection{Classical phase space}

For the Schwarzschild black hole solution, the line element in spherical coordinates reads ($\theta\in [0,\pi],\ \phi \in [0,2\pi), r\geqslant 0$)
\begin{equation}\label{Schwarzschild}
ds^2 = -\big(1- \frac{r_S}{r}\big) dt^2 + \frac{1}{1- \frac{r_S}{r}}dr^2 + r^2 d\Omega^2,
\end{equation}
where $r_S = 2GM$ is the Schwarzschild radius, given by the mass $M$ of the black hole, and $d\Omega := d\theta + \sin{\theta}d\phi$. Unlike the static exterior region, the Schwarzschild solution in the interior region ($0 \leq r \leq r_S$) of the black hole is dynamical, as the Killing vector field $\partial_t$ is space-like while the vector field $\partial_r$ is time-like inside the black hole. As such, in the interior region, the spacetime metric can be re-cast into the form of a Kantowski-Sachs metric \cite{kantowskiSachs}, where spacetime is homogeneous and anisotropic. We therefore can treat the interior region as a dynamical system with a finite dimensional phase space, with non-trivial equations of motion to be solved. The Kantowski-Sachs line element is given as
\begin{equation}
    ds^2 = -N^2(t) dt^2 + f^2(t) dr^2 + g^2(t) d\Omega^2,
    \label{eq:KSmetric}
\end{equation}
where Eintein's equations introduce evolution equations for the functions $N$, $f$ and $g$. The analysis of the dynamics can be performed using the Hamiltonian formulation of general relativity based on Ashtekar-Barbero variables, consisting of the densitized triad $E^a_i$ and the connection $A^i_a$, which satisfy the Poisson brackets
\begin{equation}
    \{A^i_a(x), E^b_j(y)\} = \kappa \beta \delta_a^b \delta_j^i \delta^3(x,y).
\end{equation}
where $\beta$ is the Immirzi-Barbero parameter. For the Kantowski-Sachs line element given in eq.\,\eqref{eq:KSmetric}, we choose these variables to be given as
\begin{equation}
    E^r_r = \text{sgn}(f) g^2\sin{\theta} , \quad E^{\theta}_{\theta} = |f| g\sin{\theta} , \quad E^{\phi}_{\phi} = |f| g ,
\end{equation}
\begin{equation}
    A_r^r = \beta N^{-1} \Dot{f} , \quad A_{\theta}^{\theta} = \beta N^{-1} \Dot{g} , 
    \quad A_{\phi}^{\phi} = \beta N^{-1} \Dot{g}\sin{\theta} , \quad A^r_{\phi} = \cos{\theta},
\end{equation}
Based on the symplectic structure of the general phase space, one can identify two pairs of canonical variables for the reduced Kantowski-Sachs phase space, which we define as
\begin{equation}\label{can.var.}
\begin{aligned}
   a :=&\, R_0 \beta N^{-1}\dot{f},& \quad p_a :=&\, \text{sgn}(f) g^2, \\
   b :=&\, \beta N^{-1} \dot{g},& \quad p_b :=& \, R_0 2 |f| g ,
\end{aligned}
\end{equation}
where the factor $R_0$ comes from the symplectic form and it originates from the integration over a compact region of the spatial hypersurface, a.k.a. the fiducial cell, in which the radial coordinate used in eq.\,\eqref{eq:KSmetric} is restricted to $0\leq r \leq R_0$.
The dependence on $R_0$ is chosen to be absorbed in $a$ and $p_b$, which avoids a dependence of the Poisson bracket on the fiducial cell, namely
\begin{equation}
    \{a, p_a\} = \frac{\kappa \beta}{4 \pi} =  \{b, p_b\}.
\end{equation}
Under a re-scaling of the fiducial cell $R_0 \rightarrow \xi R_0$, the new variables scale as $ a \rightarrow \xi a,\  p_b \rightarrow \xi p_b$, while $p_a$ and $b$ remain unchanged.

It then follows that the line element in eq.\,\eqref{eq:KSmetric} can be expressed in terms of the new variables as
\begin{equation}
    ds^2 = -N(t)^2 dt^2 + \frac{p_b(t)^2}{4 R_0^2 |p_a(t)|}dr^2 + |p_a(t)|d\Omega^2.
    \label{eq:expansion_lineEl}
\end{equation}

The Hamiltonian analysis of the system leads to the classical Hamiltonian
\begin{equation}
    H_{\text{cl}}(a,b, p_a, p_b) =\,
   - \frac{2 \pi N \text{sgn}(p_b)}{\kappa \beta^2 \sqrt{|p_a|}} \Big(4 a b p_a +  (\beta^2 + b^2) p_b\Big),
   \label{eq:Hclnv}
\end{equation}
and the equations of motion for the variables $(a,b,p_a,p_b)$ can be solved analytically for a nicely chosen lapse function $N$. For instance, for $N=\text{sgn}(p_b)\sqrt{|p_a|}$ one obtains
\begin{align} \label{class-analytical-sol}
\begin{array}{lll}
a(t) = a^0 \cos\left(\dfrac{t-t_0}{2}\right)^{-4 \text{sgn}(p_a^0)} & , & \qquad p_a(t) = p_a^0 \cos\left(\dfrac{t-t_0}{2}\right)^{4 \text{sgn}(p_a^0)},
\\
\\
b(t) = -\beta \tan\left(\dfrac{t-t_0}{2}\right) & , & \qquad p_b(t) = \dfrac{2}{\beta} a^0 p_a^0\sin\left(t-t_0\right),
\end{array}
\end{align}
where $a^0 := a(t_0)$ and $p_a^0 := p_a(t_0)$ are integration constants. The comparison with the Schwarzschild line element in eq.\,\eqref{Schwarzschild} indicates that $b(t_0) = 0 = p_b(t_0)$ identify $t = t_0$ with the black hole horizon, as it is characterized by the vanishing of the Killing vector field, while $p_a^0 = r_S^2$. The singularity is identified by the condition $p_a=0$, occurring at the point $t= t_0 \pm \frac{\pi}{2}$ ($+$ or $-$ if the lapse $N$ is taken to be positive or negative respectively) and it translates into a diverging spacetime curvature, e.g. the divergence of the Kretschmann scalar $K := R^{\alpha \mu \beta \nu}  R_{\alpha \mu \beta \nu} = 12\frac{p_a^0}{p_a^3}$.

\subsection{Effective Schwarzschild black hole models}

A lot of effort has been directed towards defining a quantum model for the Schwarzschild black hole interior in the context of LQG  \cite{Qbh_Ashtekar, Qbh_Modesto1, Qbh_Modesto2, Qbh_Pullin1, Qbh_Pullin2, Qbh_Pullin3, Qbh_Pullin4}. On the other hand, the success of loop effective models in capturing the quantum corrections induced by LQC dynamics has inspired the construction and study of effective black hole models \cite{CS, AOS, BMM, 41mubarR0independent, 42mubarR0dependent,mu0}, which are not necessarily derived from a quantum model.
As a consequence, several ambiguities exist in the procedure of constructing the effective dynamics for the black hole interior, and as outlined below, these ambiguities influence the resulting phenomenology. 
By probing the effects of different choices made in the construction of effective models, one can ultimately hope to gain some guidance for completing the construction of a quantum theory for black holes. 

From a physical viewpoint, there exist some properties that an effective model of the black hole interior should fulfill. First, the singularity encountered at the center of the black hole should be lifted, i.e.\ there should exist a universal upper bound on the spacetime curvature. This means that the regime in which quantum effects become important is marked by high curvatures. 
Secondly, any effective theory should agree with the classical one in the low curvature regime; namely, far away from the singularity, where curvature becomes small, quantum effects should fade. This also implies that, in a physically sensible model, the upper curvature bound cannot be arbitrarily small, otherwise quantum effects would occur in the classical regime. 
Finally, quantities such as observables and physical scales should not depend on the size of the fiducial cell $R_0$, as it is an unphysical auxiliary structure.

Several loop effective models of the black hole interior discussed in the literature \cite{CS, AOS, BMM, 41mubarR0independent, 42mubarR0dependent,mu0} successfully manage to satisfy the properties mentioned above. Additionally, they predict the existence of a second Killing horizon, a ``white hole''-like horizon, to which one can associate a mass $M_{\rm wh}$ which may depend on the black hole mass $M_{\rm bh}$. When that is the case, the form of the relation between $M_{\rm wh}$ and $M_{\rm bh}$ depends strongly on the form of the effective Hamiltonian and on the choice of regulators.

Effective models are commonly based on the idea of introducing a modified Hamiltonian for the system, the effective Hamiltonian. Such a Hamiltonian is often obtained by altering the dependence of the classical Hamiltonian (eq.\,\eqref{eq:Hclnv}) on the phase variables through a regularization of the variables associated to the Ashtekar connection \cite{CS,AOS,BMM}. 
For instance, if the Ashtekar connection is regularized as in \cite{CS,AOS,BMM}, this would consist of the replacement
\begin{equation}
    a\rightarrow \frac{\sin{(\mu_a a)}}{\mu_a}, \quad  b\rightarrow \frac{\sin{(\mu_b b)}}{\mu_b},
    \label{eq:poly}
\end{equation}
in  eq.\,\eqref{can.var.}, where $\mu_a$ and $\mu_b$ are regularization parameters. It is straightforward to see that in the limit $|\mu_a a|\ll 1$ and $|\mu_b b|\ll 1$ the classical Hamiltonian is recovered. Hence, independently of the choice of the regularization parameters $\mu$, it is always required that in the classical regime the conditions $|\mu_a a|\ll 1$ and $|\mu_b b|\ll 1$ are satisfied.

The choices of regularization parameters $\mu$ can be categorized into different regularization schemes. One can roughly distinguish between the $\mu_o$-scheme, the generalized $\mu_o$-scheme (e.g.\ in \cite{CS,AOS}), and the $\bar{\mu}$ scheme (e.g.\ in \cite{BMM}). In the $\mu_o$-scheme, the regularization parameters are set to be a constant on the whole phase space. In the generalized $\mu_o$-scheme the regularization parameters are constant along each dynamical phase space trajectory, and therefore depend on the initial conditions characterizing each trajectory (such as the mass of the black hole). Finally, in the $\bar{\mu}$-scheme the regularization parameters are taken to be functions of the phase space variables.
The latter choice is motivated by the success of $\bar{\mu}$-schemes in homogeneous and isotropic quantum cosmology, where the phase space dependence cures issues arising in the $\mu_o$-scheme, which mainly concern the dependence of physical quantities on the choice of the fiducial cell \cite{LQC1improvedDyn, LQC1uniqueMubar}.

A more involved approach to obtain the effective Hamiltonian for the black hole interior was introduced in \cite{mu0} and was studied in a $\mu_o$-scheme. This approach relies on the regularization methods used in LQG to construct the Hamiltonian operator, in particular on the so called Thiemann identities \cite{thiemannQSD1, thiemannBook}. This method to regularize the Hamiltonian was already applied in the context of homogeneous and isotropic cosmology, both in the implementation of the quantum dynamics \cite{deSitter1, deSitter2} as well as the effective one \cite{andreKlaus}. The results describe an evolution which differs significantly from previously studied models, where one proceeds with a regularization of the reduced Hamiltonian without using Thiemann identities. This observation motivated the work in \cite{mu0} in order to evaluate possible differences between the two methods. It turns out that one recovers the same qualitative features as the other effective models for the Schwarzschild black hole interior, namely the singularity resolution and a second horizon. 
However, for the same initial condition for the variable $a$, the relations between the black hole and white hole masses differ.

Nevertheless, as mentioned above, the $\mu_o$-scheme is known to introduce non-physical effects to the models; hence the need to study the dynamics generated by effective Hamiltonians obtained using Thiemann identities in the $\bar{\mu}$-scheme, including what choices of the regularization parameters are viable and evaluate how they may differ. These are the questions we investigate in the present article, as detailed below.

\section{Effective models with modified dynamics}

\subsection{Modified Hamiltonian}\label{sec:newM_mu}

In the model for the Schwarzschild black hole interior presented in this article, the effective Hamiltonian is constructed using Thiemann identities as was already done in \cite{mu0}, but with the use of phase space functions as regularization parameters ($\bar{\mu}$-scheme). 

One may recall that the general classical form of the Hamiltonian of general relativity in terms of Ashtekar-Barbero variables \cite{thiemannBook} is given by the sum of the Euclidean part $H_E$ and the Lorentzian part $H_L$:
\begin{equation}
    \begin{aligned}
    H[N] &= H_E[N] + H_L[N],\\
    H_E[N]&=  \int_{\Sigma_0} d^3x \, \frac{N}{2 \kappa}\frac{\epsilon_{ijk}E^a_j E^b_k}{\sqrt{q}}F^i_{ab},\\
    H_L[N]& = - \int_{\Sigma_0} d^3x \, \frac{N}{2 \kappa}(1+\beta^2)\frac{\epsilon_{ijk}E^a_j E^b_k}{\sqrt{q}}\epsilon_{imn}K^m_aK^n_b,
    \end{aligned}
    \label{eq:HAB}
\end{equation}
where $K_a^i$ is the extrinsic curvature one-form, and the integration has been limited to a compact region $\Sigma_0$ ($0\leq r\leq R_0$) of the spatial hypersurface $\Sigma$.

Most previously studied models for the black hole interior follow the ``reduce then regularize'' approach where one first carries out a symmetry reduction to the Kantowski-Sachs phase space, on which the Hamiltonian takes a relatively simple form, and then regularizes. The alternative approach, which we follow here, is to instead regularize the Hamiltonian as a functional on the full phase space first and then reduce to the Kantowski-Sachs setting. 

Using the Thiemann identities, namely
\begin{equation}
\begin{aligned}
 \text{sgn}(\det E)\frac{\epsilon^{ijk}E^a_j E^b_k}{\sqrt{q}}&= \frac{2}{\kappa \beta} \epsilon^{abc}\{ A^i_c, V\},\\
     K^i_a&= \frac{1}{\kappa \beta}\{A^i_a, K\},
     \label{eq:thiemannKey}
\end{aligned}
\end{equation}
where $V:=\int_{\Sigma_0} d^3x \sqrt{|\det E|}$ is the volume of the compact spatial region $\Sigma_0$, and $K:=\int_{\Sigma_0} d^3x K_a^i E_i^a = \{H_E[1], V\}/\beta^2$ is the extrinsic curvature scalar, the functionals $H_E$ and $H_L$ can be re-expressed as
\begin{equation}\label{eq:HeHl}
\begin{aligned}
    H_E[N] &= \frac{\text{sgn}(\det E)}{\kappa^2 \beta} \int_{\Sigma_0} d^3x \ N \epsilon^{abc}F^i_{ab}\{A^i_c, V\},\\
    H_L[N] &= -\frac{\text{sgn}(\det E)}{\kappa^4\beta^7} (1+\beta^2)\int_{\Sigma_0} d^3x \ N \epsilon^{abc}\epsilon_{ijk}\{A^i_a, \{H_E[1], V\}\}\{A^j_b,\{H_E[1], V\}\}\{A^k_c,V\}.
\end{aligned}
\end{equation}
One can then regularize the terms in the final expression as it is done in LQG \cite{thiemannQSD1}, using holonomies and fluxes.
In the black hole setting, one considers holonomies along coordinates $r, \theta$ and $\phi$, which are generally given by
\begin{equation}
\begin{aligned}
   h_{r}^{\mu_a} =& \exp(-\mu_a a \tau_1) = \cos  \Big(\frac{\mu_a a }{2}\Big) - 2\tau_1\sin\Big(\frac{\mu_a a }{2}\Big),\\
    h_{\theta}^{\mu_b} = &\exp(-\mu_b b \tau_2) = \cos\Big(\frac{\mu_b b }{2}\Big) - 2\tau_2\sin\Big(\frac{\mu_b b }{2}\Big),\\
    h_{\phi}^{\mu_b}  =& \exp(- \mu_b b \tau_3\sin{\theta} + \mu_b\tau_1\cos{\theta}),\\
\end{aligned}
\label{eq:holGen}
\end{equation}
where the edge in the $r$ direction has a coordinate length $\mu_a$, while the edges in the ${\theta}$ and ${\phi}$ directions have the same coordinate lengths $\mu_b$, and $\tau_i$ are the $SU(2)$ group generators in the $\frac{1}{2}$ representation, given by the Pauli matrices $\tau_i = -\frac{i}{2} \sigma_i$.
As discussed in \cite{mu0}, the general form of $h^{\mu_b}_{\phi}$ in eq.\,\eqref{eq:holGen} leads to very lengthy expressions for $H_E$ and $H_L$, where the integration over $\theta$ can be carried out only numerically, and the resulting equations of motion are too involved to be solved numerically. For this reason one has to simplify the choice of the holonomies used for the regularization a bit further, which is achieved by choosing a subset of the holonomies along the $\phi$ direction, identified by a fixed value of the $\theta$ coordinate. In our case we take $\theta = \frac{\pi}{2}$, which induces the most simplified form of the effective Hamiltonian that allows solving the effective equations of motion numerically\footnote{It is unclear at this point whether a different choice of $\theta$ would lead to qualitatively different phenomenological results with respect to the $\theta = \frac{\pi}{2}$ case, as we were unable to solve the equations of motion induced by the $\theta$ dependent effective Hamiltonian. Furthermore, if the integration over the coordinate $\theta$ was possible, this may introduce additional holonomy corrections which could significantly alter the resulting effective dynamics and the consequent phenomenology.}. Hence the set of holonomies we use is
\begin{equation}
\begin{aligned}
    h_{r}^{\mu_a} & = \cos\Big(\frac{\mu_a a}{2}\Big) - 2\tau_1 \sin\Big(\frac{\mu_a a}{2}\Big),\\
      h_{\theta}^{\mu_b} & = \cos\Big(\frac{\mu_b b}{2}\Big) - 2\tau_2 \sin\Big(\frac{\mu_b b}{2}\Big),\\
       h_{\phi}^{\mu_b}\big|_{\theta = \frac{\pi}{2}} & = \cos\Big(\frac{\mu_b b}{2}\Big) - 2\tau_3 \sin\Big(\frac{\mu_b b}{2}\Big).
    \end{aligned}
    \label{eq:holSimple}
\end{equation}
The drawback of this simplified choice of holonomies in the $\phi$ direction is that one is unable to properly approximate certain components of the curvature $F_{ab}^i$ of the Ashtekar connection via a simple product of the holonomies given in eq.\,\eqref{eq:holSimple}. This is because the term stemming from the $A^r_{\phi}$ component of the Ashtekar connection vanishes for $\theta = \frac{\pi}{2}$. 
Therefore, when constructing the effective Euclidean part of the Hamiltonain in eq.\,\eqref{eq:HeHl} and approximating the curvature $F^i_{ab}$ by holonomies, one must correct the regularized expression by adding the missing term which is proportional to the first Thiemann identity in eq.\,\eqref{eq:thiemannKey}, as was done in \cite{CS}.
Before diving into the details of the implementation of the $\bar{\mu}$-scheme, let us for the moment assume that we are working in the $\mu_o$-scheme. The complete calculation following the abovementioned procedure turns out to give the same result as when one first reduces $H_E$ to the Kantowski-Sachs phase space, then regularizes using eq.\,\eqref{eq:poly}, that is
\begin{equation}
    H_{E, \text{eff}}^o[N] = \frac{ 4 \pi N \text{sgn}(p_b)}{\kappa \sqrt{|p_a|}} \Bigg(2p_a\frac{\sin{(\mu_a a)}\sin{(\mu_b b)}}{\mu_a \mu_b} + \frac{p_b}{2}\Big(\frac{\sin{(\mu_b b)}^2}{\mu_b^2} -1\Big)\Bigg),
    \label{eq:HEeff}
\end{equation}
where the superscript ``$o$'' denotes the $\mu_o$-scheme.
Once we have the effective Euclidean part of the Hamiltonian $H_{E,\text{eff}}^o$, the effective Lorentzian part $H_{L,\text{eff}}^o$ is calculated using the regularized Thiemann identities. This allows to obtain an effective Hamiltonian for the black hole interior using the methods employed in LQG, and which is still concise enough to produce equations of motion that can be solved numerically. For a phase space independent $\mu=\mu_o$, the regularized Thiemann identities are given by
\begin{align}
  \tau_i \{ A^i_c, V\}
   &\approx \frac{\Lambda_c}{\mu_c}  \,  h_c^{\mu_c} \{h_c^{\mu_c\dagger}, V\}, 
  \label{eq:thiemann1}\\
  \tau_i  \{A^i_c, \{H_E[1], V\}\}
   &\approx  \frac{\Lambda_c}{\mu_c} h_c^{\mu_c}\{h_c^{\mu_c \dagger},\{H^o_{E,\text{eff}}[1],V\}\},
  \label{eq:thiemann2}
\end{align}
with $V = 2\pi |p_b| \sqrt{|p_a|}$ being the volume of the compact spatial slice $\Sigma_0$, and $\Lambda =\big(R_0^{-1}, 1, \sin{\theta}\big)^T$.
The factor $R_0^{-1}$ in $\Lambda_1$ assumes that $\mu_a$ is independent of the fiducial cell. The calculations then lead to the effective Lorentzian part of the Hamiltonian being
\begin{align}
\begin{aligned}
H_{L,\text{eff}}^o[N] =  & - 4 \pi  \frac{1+\beta^2}{\kappa \beta^2} \frac{N \text{sgn}(p_b)}{\sqrt{|p_a|}}\big(\cos(\mu_a a) + \cos(\mu_b b)\big) \sin(\mu_b b) \times\\
   & \ \Big[p_a 
   \frac{ \sin(\mu_a a)  \cos(\mu_b b)}{\mu_a \mu_b} + \frac{p_b  \big(\cos(\mu_a a) + \cos(\mu_b b)\big) \sin(\mu_b b)}{8\mu_b^2}
     \Big].\label{eq:HLeff}
\end{aligned}
\end{align}
It follows that the effective Hamiltonian reads
\begin{equation}
    \begin{aligned}
     H_{\text{eff}}^o [N] & = \frac{4\pi}{\kappa} \frac{N\text{sgn}(p_b)}{\sqrt{|p_a|}}
  \Bigg( 2 p_a
     \frac{ \sin(\mu_a a)\sin(\mu_b b) }{\mu_a \mu_b} +
     \frac{p_b}{2}\Big(
      \frac{\sin(\mu_b b)^2 }{\mu_b^2}  - 1\Big)   \\
      & \quad - \frac{1+\beta^2}{\beta^2}(\cos(\mu_a a) + \cos(\mu_b b)) \sin(\mu_b b) \times\\
   &\quad\quad \Big[ p_a
   \frac{\sin(\mu_a a)  \cos(\mu_b b)}{\mu_a \mu_b} + \frac{p_b  \big(\cos(\mu_a a) + \cos(\mu_b b)\big) \sin(\mu_b b)}{8\mu_b^2}
    \Big] \Bigg).
    \end{aligned}
    \label{eq:h2effmu0}
\end{equation}
Note that the above expression gives the classical Hamiltonian in the limit $|\mu_a a| \rightarrow 0$ and $|\mu_b b| \rightarrow 0 $.
At this point it shall again be pointed out that eq.\,\eqref{eq:thiemann1} and eq.\,\eqref{eq:thiemann2} are only valid in the $\mu_o$-scheme.
Therefore, when introducing a phase space dependent regularization parameter $\bar{\mu}$, its form must be considered carefully in order to ensure that the correct classical limit is recovered.

In the following, we analyze some choices of phase space dependent regularization parameters and their compatibility with Thiemann identities in order to construct an admissible effective Hamiltonian following the ``regularize then reduce'' approach.
We first consider the following rather general monomial form of $\bar{\mu}$, which depends on all four phase space variables 
\begin{equation}
\bar \mu_a \propto |a|^q |b|^k |p_a|^{-n} |p_b|^{-v}, \quad \bar \mu_b \propto |a|^u |b|^l |p_a|^{-w} |p_b|^{-m},
\label{eq:mubargeneral}
\end{equation}
and where the values of the exponents can be chosen independently for $\bar{\mu}_a$ and $\bar{\mu}_b$.
Naturally, the values of these parameters are restricted due to certain physical considerations:
the horizon of large black holes is far from the black hole singularity and the classical limit where $|\bar{\mu}_a a| \ll 1$ and $|\bar{\mu}_b b | \ll 1 $ must therefore apply.
Despite $p_b$ vanishing at the black hole horizon, the classical solutions (eq.\,\eqref{class-analytical-sol}) show that the combination $b^0 (p_b^0)^{-1}$ is not singular. To ensure $|\bar{\mu}_a a| \ll 1$ and $|\bar{\mu}_b b | \ll 1 $ at the horizon it is therefore required that $k\geqslant v$  and  $l+1 \geqslant m$. 

It is now to be verified if the regularized Thiemann identities in eq.\,\eqref{eq:thiemann1} and eq.\,\eqref{eq:thiemann2} hold for a choice of regularization parameters as given above. It turns out that it is possible to identify an adapted version of the regularized identities, which recover the correct classical limit, if one imposes certain restrictions on the exponents in eq.\,\eqref{eq:mubargeneral}.

The comparison between the right-hand side of eq.\,\eqref{eq:thiemann1} calculated in the $\bar \mu$-scheme and the classical expression on the left-hand side leads to
\begin{equation}\label{1stRestrictions}
k = u = 0.
\end{equation}
Furthermore, unlike in the $\mu_o$ case, eq.\,\eqref{eq:thiemann1} requires an additional overall factor in order to hold in the considered $\bar \mu$-scheme.
Namely, the modified version of the first Thiemann identity eq.\,\eqref{eq:thiemann1} is given by
\begin{equation}
\tau_i \{ A^i_c, V\}
   = \eta_c\, \frac{2}{\kappa \beta} \frac{\Lambda_c}{\mu_c} \,  \epsilon^{abc} \,  h_c^{\mu_c} \{h_c^{\mu_c\dagger}, V\},
\label{eq:thiemann1adapted}
\end{equation}
where $\eta_c$ corresponds to the factor $(1+ q)^{-1}$ for $\bar \mu_c=\bar\mu_a$ and $(1+l)^{-1}$ for $\bar \mu_c=\bar\mu_b$. 
The special case $q=-1$ can be realized by taking the limit $q\rightarrow -1$ after having obtained a general expression in terms of $q$ for the right-hand side of  eq.\,\eqref{eq:thiemann1adapted}. The same applies to $l=-1$.
Using eq.\,\eqref{eq:thiemann1adapted}, the Euclidean part of the effective Hamiltonian in the $\bar \mu$-scheme is given by simply replacing the $\mu$ parameters in the expression of $H_{E, \text{eff}}^o[N]$ by the corresponding $\bar \mu$ parameters.
This leads to an effective Euclidean part $H_{E,\text{eff}}$ of the form
\begin{align}
 H_{E, \text{eff}}[N] = \frac{ 4 \pi }{\kappa} \frac{N \text{sgn}(p_b)}{\sqrt{|p_a|}} \Bigg(2 p_a \frac{\sin{(\bar\mu_a a )}\sin{(\bar\mu_b b )}}{\bar\mu_a \bar\mu_b} + \frac{p_b}{2}\Big(\frac{\sin{(\bar\mu_b b )}^2}{\bar\mu_b^2} -1\Big)\Bigg).
    \label{eq:HEeffbar}
\end{align}

The calculation of the right-hand side of the second Thiemann identity eq.\,\eqref{eq:thiemann2} in the $\bar \mu$-scheme and its comparison with the classical limit, with $H_{E, \rm eff}$ as given above, allows to identify additional conditions on the exponents of the regularization parameters in eq.\,\eqref{eq:mubargeneral}, namely
\begin{equation}
\begin{aligned}
n\, u = (1+q) w, \quad \ k\, m   = (1 + l) v .
\end{aligned}
\end{equation}
Together with eq.\,\eqref{1stRestrictions}, the final set of conditions obtained from Thiemann identities and the semiclassical limit is
\begin{align}\label{mubarRestrictions}
 k=u=0, \qquad  (1+q) w = (1 + l) v = 0, \qquad v\leqslant 0, \qquad 
 m \leqslant l+1.
\end{align}
Hence, the general monomials which can be considered for the $\bar \mu$ parameters are 
\begin{equation}
\bar \mu_a = |a|^q |p_a|^{-n}|p_b|^{-v} ,\ \bar \mu_b = |b|^l|p_a|^{-w} |p_b|^{-m}.
\label{eq:mubarRestricted}
\end{equation}
The calculation of the regularized expression in eq.\,\eqref{eq:thiemann2} gives
\begin{equation}\label{KiaEff}
\begin{aligned}
  \frac{1}{\kappa \beta^3 \bar{\mu}_a} h_r^{\bar{\mu}_a}\{h_r^{\bar{\mu}_a \dagger},\{H_{E,\text{eff}}[1],V\}\}  = &
 \frac{ \text{sgn}(p_b)}{2 \beta \,b^2 \bar\mu_a \bar\mu_b^2}  \Bigg(2 b^2 (l+1) \bar\mu_b^2 (n+q+1) \sin (\bar\mu_a a ) \cos (\bar\mu_b b)\\
& +\sin (\bar\mu_b b) \Big[-\bar\mu_b b \sin (\bar\mu_a a ) \big(2 l (n+q+1)+q v\big)\\
&+a b \bar\mu_a \bar\mu_b (q+1) v \cos (\bar\mu_a a )  +\bar\mu_a a  v \sin (\bar\mu_b b)\Big]\Bigg)\tau_1,\\
\frac{1}{\kappa \beta^3 \bar{\mu}_b} h_{\theta}^{\bar{\mu}_b}\{h_{\theta}^{\bar{\mu}_b \dagger},\{H_{E,\text{eff}}[1],V\}\} = &
\frac{\text{sgn}(p_b)}{2\beta \, a b \bar\mu_a \bar\mu_b^2 } \Bigg(\sin (\bar\mu_b b) \Big[-a l \bar\mu_a (l+m+1) \sin (\bar\mu_b b)\\
& +a b \bar\mu_a \bar\mu_b (q+1) (l+m+1) \cos (\bar\mu_a a )-\bar\mu_b b \sin (\bar\mu_a a ) \big(q (l+m+1)+2 l w\big)\Big] \\
& +b (l+1) \bar\mu_b \cos (\bar\mu_b b) \Big[\bar\mu_a a  (l+m+1) \sin (\bar\mu_b b)+2 \bar\mu_b b w \sin (\bar\mu_a a )\Big]\Bigg)\tau_2,\\
\frac{1}{\kappa \beta^3 \bar{\mu}_b} h_{\phi}^{\bar{\mu}_b}\{h_{\phi}^{\bar{\mu}_b \dagger},\{H_{E,\text{eff}}[1],V\}\} = &
\frac{\text{sgn}(p_b)}{2\beta \, a b \bar\mu_a \bar\mu_b^2 } \Bigg(\sin (\bar\mu_b b) \Big[-a l \bar\mu_a (l+m+1) \sin (\bar\mu_b b)\\
& +a b \bar\mu_a \bar\mu_b (q+1) (l+m+1) \cos (\bar\mu_a a )-\bar\mu_b b \sin (\bar\mu_a a ) \big(q (l+m+1)+2 l w\big)\Big] \\
& +b (l+1) \bar\mu_b \cos (\bar\mu_b b) \Big[\bar\mu_a a  (l+m+1) \sin (\bar\mu_b b)+2 \bar\mu_b b w \sin (\bar\mu_a a )\Big]\Bigg)\tau_3.
\end{aligned}
\end{equation}
where in the last equation we use the short notation $h_{\phi}^{\bar{\mu}_b} = h_{\phi}^{\bar \mu_b}\big|_{\theta = \frac{\pi}{2}}$ as given in eq.\,\eqref{eq:holSimple}.
These expressions\footnote{Note that the expressions of the $\theta$ component and the $\phi$ component in eq.\,\eqref{KiaEff} are the same, up to the exchange of $\tau_2$ and $\tau_3$. This is a property of the choice of holonomies in eq.\,\eqref{eq:holSimple} and holds for arbitrary $\bar\mu$.} do not exactly reduce to the extrinsic curvature $1$-form $\tau_i K^i_c $ in the classical limit, but they introduce overall factors in the results, namely
\begin{align}\label{RegExtCurv}
    \tau_i K^i_c 
   &\approx \zeta_c \frac{\Lambda_c}{\kappa \beta^3}  \frac{1}{\bar\mu_c} h_c^{\bar\mu_c}\{h_c^{\bar\mu_c \dagger},\{H_{E,\text{eff}}[1],V\}\},
\end{align}
where the factor $\zeta_c$ gives $(q+ n + v +1)^{-1}$ for $\bar\mu_c=\bar{\mu}_a$, and $(l+m+w+1)^{-1}$ for $\bar\mu_c=\bar{\mu}_b$. 
Again, to realize the cases $q+n+v =-1$ and $l+m+w =-1$, the limits  $q+n+v \rightarrow -1$ and $l+m+w \rightarrow -1$ can be taken after having obtained a general expression according to eq.\,\eqref{RegExtCurv}.

The Lorentzian part of the effective Hamiltonian obtained from the modified identities given in eq.\,\eqref{eq:thiemann1adapted} and eq.\,\eqref{RegExtCurv} with the choice of $\bar{\mu}_a$ and $\bar{\mu}_b$ as specified in eq.\,\eqref{eq:mubarRestricted}   recovers $H_{L}^{\rm cl}$ in the classical limit.
Explicitly, the Lorentzian part of the Hamiltonian in the ``regularize then reduce'' approach reads
\begin{equation}
\begin{aligned}
H_{L,\text{eff}} = & 
\frac{\pi  (\beta ^2+1 )}{2 \beta ^2 \kappa \, a^2 b^3 \, \bar\mu_a \bar\mu_b^4\, \sqrt{|p_a|} }\Big(l+m+w+1) \big((l+1) (n+q+1)+m (n+q+v+1)+w (n+v)\big)\Big)^{-1}\\
& \times \Bigg(a (l+m+1) \sin (\bar\mu_b b) \Big[\bar\mu_b b (q+1) \cos (\bar\mu_a a )-l \sin (\bar\mu_b b)+b (l+1) \bar\mu_b \cos (\bar\mu_b b)\Big] \\
& +\frac{1}{\bar\mu_a} \bar\mu_b b \sin (\bar\mu_a a ) \Big[2 b (l+1) \bar\mu_b w \cos (\bar\mu_b b)-\sin (\bar\mu_b b) \big(q (l+m+1)+2 l w\big)\Big]\Bigg) \\
& \times \Bigg(b^2 (l+1) \bar\mu_b \cos (\bar\mu_b b) \Big[-2 \bar\mu_b \sin (\bar\mu_a a ) \Big(4 a p_a \big((l+m+1) (n+q+1)+n w\big)+ n w b p_b  \Big) \\
& -\bar\mu_a a  p_b (l+m+1) (n+q+1) \sin (\bar\mu_b b)\Big]+\sin (\bar\mu_b b) \Big[\bar\mu_b b \sin (\bar\mu_a a ) \\
& \times \Big(4 a p_a \big(2 l^2 (n+q+1)+2 l \big((m+1) (n+q+1)+n w\big)+v (m q-w)\big) \\
& +b p_b \big(q (l+m+1) (n+q+1)+2 w (l n-v)+m q v\big)\Big) \\
& -a b \bar\mu_a \bar\mu_b (q+1) \cos (\bar\mu_a a ) \Big(4 a m p_a v+b p_b \big((l+m+1) (n+q+1)+m v\big)\Big)\\
&
+\bar\mu_a a  \sin (\bar\mu_b b) \Big(b p_b \big(l (l+m+1) (n+q+1)-m v\big)-4 a p_a v (m+w)\Big)\Big]\Bigg).
\end{aligned}
\end{equation}
The analysis and discussion of the regularization parameters above demonstrates that, in the context of the $\bar \mu$-scheme, the viability of Thiemann identities in the construction of the effective Hamiltonian, or even the quantum Hamiltonian operator in quantum reduced models, is not always guaranteed. Often, one must adjust the final expressions in order to recover the correct classical limit. 

This concludes the general construction of the Euclidean and Lorentzian parts of the effective Hamiltonian. In the remainder of the article, we focus on the specific choice of regularization parameters in which $\bar{\mu}_a$ depends only on $p_a$ and $\bar{\mu}_b$ depends only on $p_b$ (i.e. we set $s=v=l=w=0$). This would allow to make a closer comparisons with previously studied models in the literature \cite{CS, AOS, BMM}. Also, recall that under a resizing of the fiducial cell $R_0 \rightarrow \xi R_0$, $a$ scales as $a\rightarrow \xi a$ and $b$ remains unaffected $b\rightarrow b$. To protect the effective models from a dependence on $R_0$ as it was observed in \cite{R0}, the regularization parameters must be chosen such that $\bar{\mu}_a \propto R_0^{-1}$ and $\bar \mu_b$ is independent of $R_0$. Explicitly, our choice of regularization parameters, with the required scaling behavior imposed, is given by
\begin{equation}
    \bar{\mu}_a = \frac{j_a}{R_0} |p_a|^{-n},\quad  \bar{\mu}_b = j_b R_0^m |p_b|^{-m},
    \label{eq:decoupledMubar}
\end{equation}
where $j_a$ and $j_b$ are arbitrary numerical factors. 
In this case, the restrictions on the exponents in eq.\,\eqref{mubarRestrictions} reduce to imposing $m \leqslant 1$.
For this choice, and when using the same form of holonomies as in eq.\,\eqref{eq:holSimple}, the identity in eq.\,\eqref{eq:thiemann1} still holds. 
The form of the effective Hamiltonian in eq.\,\eqref{eq:h2effmu0} can then be recovered for our choice of $\bar{\mu}$, if the altered identity in eq.\,\eqref{RegExtCurv} is used instead of eq.\,\eqref{eq:thiemann2}.

The effective Lorentzian part of the Hamiltonian then takes the form 
\begin{align}\label{eq:HLeffbar}
\begin{aligned}
 H_{L,\text{eff}}[N] =  & - 4 \pi \frac{1+\beta^2}{\kappa \beta^2} \frac{N \text{sgn}(p_b)}{\sqrt{|p_a|}}
 \big(\cos(\bar\mu_a a) + \cos(\bar\mu_b b)\big) \sin(\bar\mu_b b) \times\\
   & \Big[p_a
   \frac{ \sin(\bar\mu_a a)  \cos(\bar\mu_b b)}{\bar\mu_a \bar\mu_b} + \frac{p_b  (\cos(\bar\mu_a a) + \cos(\bar\mu_b b)) \sin(\bar\mu_b b)}{8\bar\mu_b^2}
     \Big]
\end{aligned}
\end{align}
and together with eq.\,\eqref{eq:HEeffbar}, we obtain the effective Hamiltonian $H_{\text{eff}}$ for the choice of $\bar\mu$ given in eq.\,\eqref{eq:decoupledMubar}, namely
\begin{align}\label{eq:Heffbar}
\begin{aligned}
H_{\text{eff}} [N] & = \frac{4\pi}{\kappa}\frac{N \text{sgn}(p_b)}{\sqrt{|p_a|}}
  \Bigg( 2 p_a
    \frac{ \sin(\bar\mu_a a)\sin(\bar\mu_b b) }{\bar\mu_a \bar\mu_b} +
    \frac{p_b}{2}\Big(
    \frac{\sin(\bar\mu_b b)^2 }{\bar\mu_b^2}  - 1\Big)   \\
    & \quad - \frac{1+\beta^2}{\beta^2}(\cos(\bar\mu_a a) + \cos(\bar\mu_b b)) \sin(\bar\mu_b b) \times\\
    &\quad\quad \Big[ p_a
    \frac{ \sin(\bar\mu_a a)  \cos(\bar\mu_b b)}{\bar\mu_a \bar\mu_b} + \frac{p_b  (\cos(\bar\mu_a a) + \cos(\bar\mu_b b)) \sin(\bar\mu_b b)}{8\bar\mu_b^2}
    \Big] 
   \Bigg).
\end{aligned}
\end{align}
This is the Hamiltonian which defines the effective model we study in the remainder of the article. Note that, as in the case of the effective Euclidean part, both $H_{L,\text{eff}}$ and $H_{\text{eff}}$ can be respectively obtained from $H_{L,\text{eff}}^o$ and $H_{\text{eff}}^o$ by simply replacing the $\mu$ parameters by the corresponding $\bar\mu$ parameters. This is a particular feature of the choice of $\bar\mu$ parameters in eq.\,\eqref{eq:decoupledMubar}.
As discussed earlier, the new model is set up in such a way that the value of $R_0$ does not influence the value of physically meaningful parameters. Thus, the size of the fiducial cell will hereafter be set to $R_0 = 1$.

\subsection{Numerical analysis of the model}\label{sec:numAnalysis}

This section examines the phenomenology of the new model for the Schwarzschild black hole interior as given by $H_{\text{eff}}$ in eq.\,\eqref{eq:Heffbar}. 
We start by determining the initial conditions required for the numerical solving in the limit of large black holes, which requires establishing an expression for the initial value of $a$. 
Then we study the dynamics of the model and discuss the phenomenological properties, such as the singularity resolution and the presence of a second Killing horizon, in comparison to some models present in the literature. We conclude with further comments on the choice of regularization parameters.

\subsubsection{Initial conditions}
\label{sec:pheno_init}

In order to solve the equations of motion resulting from the new Hamiltonian $H_{\text{eff}}$ numerically, all four initial conditions ($ p_a^0, p_b^0, a^0, b^0$) must be determined. 
In the classical Hamiltonian analysis of the Kantowski-Sachs solution, the initial condition $p_b^0 = 0$ is imposed from the vanishing of the Killing vector field at the black hole horizon. As the line element remains unchanged, $p_b = 0$ still identifies the black hole horizon in the effective theory. 
Since the effective model cannot be solved analytically, the initial condition for $p_a$ in the effective theory is approximated in the classical limit by the classical initial value $p_a^0 = 4 M_{\rm bh}^2$, and the numerical analysis of $H_{\text{eff}}$ is restricted to black holes for which the classical limit conditions $| \bar\mu_a a |\ll 1$ and $| \bar \mu_b b| \ll 1$ are satisfied at the horizon.
This will be the case for large black holes, whose horizons are far from the singularity and therefore lie in the low curvature regime in which quantum effects should arguably become negligible. 

Next, to obtain the classically undetermined initial condition for $a$, we proceed analytically by deriving the solution to the equation of motion for $\bar{\mu}_a(t) a(t)=:\bar a(t)$, with a particular choice of lapse. We then impose classicality close to the black hole horizon. The classicality condition on $\bar a$ is translated into a condition on the Kretschmann scalar $K$ at the horizon, allowing to establish the dependence of $\bar a^0$, and equivalently $a^0$, on the mass of the black hole. 

The following, rather complicated, choice of lapse\footnote{Note that this lapse becomes singular during the evolution, which is why it cannot be used to numerically solve the equations of motion up to the second horizon.}
\begin{equation}
\begin{aligned}
 N =  -\frac{\beta\ \text{sgn}(p_b) \sqrt{|p_a|}\ \bar\mu_b} {\sin(\bar \mu_b b) \Big(2 \beta^2 - (1+\beta^2)(\cos(\bar a) + \cos(\bar \mu_b b)) \cos(\bar \mu_b b) \Big)}
\end{aligned}
\label{eq:lapseAnalytic}
\end{equation}
leads to a simple form of the equation of motion for $\bar a(t)$, namely 
\begin{align}\label{EoMa}
  \{\bar a, H_{\text{eff}}\} = - (n+1) \sin(\bar a).
\end{align}
The case of $n=-1$ leads to $\bar a(t) = const.$, which consequently does not allow to relate the classicality condition in the vicinity of the black hole horizon to any physical quantity. Hence, in this case, the initial condition $\bar a_0$ remains a free parameter which must satisfy $\bar a_0 \ll 1$.
Otherwise, the solution to the equation of motion is given by
\begin{equation}
\bar a(t) = 2 \text{arccot} \Big(\exp\Big[\mathcal{C} + (n+1)( t-t_0)\Big]\Big),
\end{equation}
where $\mathcal{C}$ is a constant to be determined, and $t=t_0$ corresponds to the black hole horizon. 
The classical regime $|\bar a|\ll 1$ is reached when the argument of the arccotangent is large, which implies $\exp(\mathcal{C})\gg 1$.
Furthermore, requiring that the system is still in the classical regime in a small neighborhood of the horizon implies 
\begin{equation}
    \exp\Big(\mathcal{C} + (n+1)( t-t_0)\Big) \gg 1
    \label{eq:classCond}
\end{equation}
for $t = t_0 + \epsilon$ with $\epsilon\ll 1$.
The classicality condition in eq.\,\eqref{eq:classCond} can be reformulated in terms of $p_a$: 
first we 
consider the classical limit of the system, where the lapse function in eq.\,\eqref{eq:lapseAnalytic} reduces to $N^{\text{cl}} = \beta\ \text{sgn}(p_b) \sqrt{|p_{a,\text{cl}}|} (2b)^{-1}$, and 
solve the classical equation of motion $\{p_{a,\text{cl}}, H_{\text{cl}}\} = p_{a,\text{cl}}$ using $N^{\text{cl}}$.
We then invert the classical solution $p_{a,\text{cl}}(t)$ to get the expression of $t(p_{a,\text{cl}})$ instead. We obtain
\begin{equation}
    p_{a,\text{cl}}(t) = p_a^0 \exp\Big(t-t_0\Big) \quad \Rightarrow \quad t(p_{a,\text{cl}}) =  t_0 + \text{ln}\Big(\frac{p_{a,\text{cl}}}{p_a^0}\Big).
\end{equation}
Inserting $t(p_{a,\text{cl}})$ into eq.\,\eqref{eq:classCond}, we get\footnote{In the classical limit, the time coordinate $t$ corresponding to the lapse $N$ in eq.\,\eqref{eq:lapseAnalytic} converges to the time coordinate $t(p_{a,\rm cl})$ corresponding to the classical lapse $N^{cl}$, therefore the substitution in \eqref{eq:classCond} is admissible as a zeroth order approximation in the classical regime, which is sufficient for the analysis here.}
\begin{equation}
     \exp(\mathcal{C})\, \Big(\frac{p_{a,\text{cl}}}{p_a^0}\Big)^{1+n} \gg 1.
     \label{eq:classCondPA}
\end{equation}
As stated above, the expression for $\exp(\mathcal{C})$ must be chosen such that eq.\,\eqref{eq:classCondPA} is satisfied in the neighborhood of the horizon.
One would prefer to write eq.\,\eqref{eq:classCondPA} as a condition on some physical quantity with respect to a universal scale that is independent of the initial conditions and of the model parameters.
There is a freedom in choosing the said quantity and we choose it to be the Kretschmann scalar $K$, based on the following physical considerations:
Being a coordinate independent measure of curvature which captures all components of the Riemann tensor, the Kretschmann scalar is arguably a suitable indicator of the classical regime. The classical approximations necessary for fixing the initial conditions would hold, if $K$ at the black hole horizon is sufficiently small.
The classical expression of $K$ for the Kantowski-Sachs spacetime is $K_{\text{cl}} = \frac{12 p_a^0}{p_{a, \text{cl}}^3}$, and inserting this into  eq.\,\eqref{eq:classCondPA} results in the following condition:

\begin{equation}
    \exp\big( \mathcal{C} \big) \Big(\frac{K_{\text{cl}}}{12} (p_a^0)^2\Big)^{-\frac{(n+1)}{3}} \gg 1 \quad \Rightarrow \quad  \left\{
    \begin{array}{l}
    K_{\text{cl}}\ll \exp\Big(\frac{3}{n+1}\mathcal{C}\Big)\, 12( p_a^0)^{-2},\ \text{if}\ n>-1 \\
    K_{\text{cl}}\gg \exp\Big(\frac{3}{n+1}\mathcal{C}\Big)\, 12( p_a^0)^{-2},\ \text{if}\ n<-1 
    \end{array}
    \right.
    \label{eq:classCondK}
\end{equation}
The case $n<-1$ suggest the presence of a lower bound on the curvature, which is in contradiction with classical evolution. Therefore, this case is to be discarded when considering the $\bar \mu$ parameters in eq.\,\eqref{eq:decoupledMubar}. On the opposite side, the case $n>-1$ suggests the presence of an upper bound on the curvature, which is a desirable feature in models resolving the singularity. 
We can therefore introduce a \emph{universal critical curvature scale} $K_{\text{crit}}$, which we define as 
\begin{align}
 K_{\text{crit}} := \exp\Big(\frac{3}{n+1}\mathcal{C}\Big)\, 12( p_a^0)^{-2},
\end{align}\label{Kcrit} 
and which is assumed to be a constant independent of model parameters that marks the regime in which quantum effects can no longer be neglected.
It then follows that
\begin{align}
 \exp\Big(\frac{3}{n+1}\mathcal{C}\Big) \propto (p_a^0)^2.
\end{align}
For all black hole masses studied in the new model, the Kretschmann scalar at the black hole horizon $K_{\text{bh}}$ must be in the low curvature regime well below $K_{\text{crit}}$, where quantum effects are negligible.
Additionally, the maximum curvature $K_{\text{max}}$ reached throughout the evolution should greatly exceed the curvature at the black hole horizon. Otherwise the singularity resolution, referred to as the \emph{bounce}, and characterized by the upper bound $K_{\text{max}}$ on the curvature, would occur in the classical regime.
The Kretschmann scalar at the bounce $K_{\text{max}}$ is not required to be below the value of $K_{\text{crit}}$, but should be of a similar order of magnitude. 
Consequently, the aspects that need to be fulfilled for physical viability are $K_{\text{bh}} / K_{\text{crit}} \ll K_{\rm max}/K_{\rm crit} \approx 1$. 
The scale $K_{\text{crit}}$ can be reformulated in terms of a ``critical mass'' $M_{\text{crit}}$, which corresponds to the mass of a black hole whose classical curvature at the horizon would be exactly $K_{\text{crit}}$,
\begin{equation}
    M_{\text{crit}}^4 := \frac{3}{4 K_{\text{crit}}}.
    \label{eq:Mcrit}
\end{equation}
For black holes with masses at the scale of $M_{\text{crit}}$, it is expected that the effective model with the given initial conditions may not be valid, and the quantum model would be required.

The value of $\mathcal{C}$ from  eq.\,\eqref{eq:classCondK} is then determined by the value of $K_{\text{crit}}$, the model parameter $n$ and the mass of the black hole, namely 
\begin{equation}
    \exp\big(\mathcal{C}\big) =  \Big(\frac{K_{\text{crit}}}{12} (p_a^0)^2\Big)^{\frac{(n+1)}{3}} = \Big(\frac{M_{\text{bh}}}{M_{\text{crit}}}\Big)^{\frac{4}{3}(n+1)}.
\end{equation}
For sufficiently large black holes, the system will be in the classical regime  at the black hole horizon and in its close proximity.
The initial condition $\bar a^0$ is then given by
\begin{equation}
 \bar a^0 = 2\ \text{arccot}\Bigg[\Big(\frac{M_{\text{bh}}}{M_{\text{crit}}}\Big)^{\frac{4}{3}(n+1)}\Bigg].
 \label{eq:amu0K}
    \end{equation}

Finally, the initial condition $b^0$ is obtained by solving the effective Hamiltonian constraint. It should be noted that $|\bar \mu_b b| \ll 1$ must also be satisfied for the classical limit to hold at the horizon. Note that in principle, one could perform 
a similar analysis as was done for the classicality condition on $\bar a$ using the classicality condition on $\bar{b} := \bar\mu_b b$ instead. 
This would lead to the determination of the integration constant for $\bar{b}$ and consequently to the determination of the initial condition for $\bar{b}$. While it may seem that the two starting points give different results, they actually should not, thanks to the fact that the Hamiltonian constraint must be satisfied at all times and that both classicality conditions $\bar a\ll 1$ and $\bar b\ll 1$ must be satisfied in the vicinity of the horizon. Hence the result must be unique. This is confirmed in the numerical analysis where, given an intial condition for $\bar{a}$ in the form of eq.\,\eqref{eq:amu0K}, there always exists a unique solution to the Hamiltonian constraint for $\bar b$ at the black hole horizon, which satisfies the classicality condition.

This concludes the question of the determination of the initial conditions for the system. 
Next, we present the results of the numerical analysis performed for a chosen range of parameters and of initial black hole masses.

\subsubsection{Numerical results}
\label{sec:pheno_num}

With the $\bar\mu$ parameters being functions on the phase space, the equations of motion for the variables $(a,b,p_a,p_b)$ get quite involved. However, for the choice of $\bar\mu$ in eq.\,\eqref{eq:decoupledMubar} and assuming\footnote{For the remainder of the article, we restrict the analysis to the case $n>-1$. The case $n=-1$ seems to produce singular equations of motion and we were unable to determine a suitable lapse function which allows the numerical solving of the system. We leave the issues surrounding this case for future studies.} $n>-1$ and $m\leq 1$ as concluded from the earlier discussions, one can introduce new canonical variables for which the equations of motion take a simpler form, namely \footnote{Note that this change of variables is a suitable choice when $m\neq-1$, which is the case for the numerical analysis presented in the following.}
\begin{equation}\label{eq:newvars}
    \begin{aligned}
         \bar a & := \bar{\mu}_a a  =a\, j_a |p_a|^{-n}, & \qquad \bar b & := \bar{\mu}_b b = b\,j_b |p_b|^{-m},\\
         \bar p_a &:= p_a\frac{|p_a|^{n}}{j_a (1+n)},&  \bar p_b & := p_b\frac{|p_b|^{m}}{j_b (1+m)}, \\
    \end{aligned}
\end{equation}
and which satisfy
\begin{align}
 \{\bar a, \bar p_a\} = \frac{\kappa \beta}{4 \pi} = \{\bar b, \bar p_b\}.
\end{align}
In terms of these new variables, the effective Hamiltonian reads
\begin{align}\label{eq:h2effmubar}
\begin{aligned}
H_{\text{eff}} [N] = & \frac{2\pi N}{\kappa \beta^2} |j_a (n+1) \bar p_a|^{-\frac{1}{2(n+1)}} \ |j_b (1+m) \bar p_b|^{-\frac{1}{m+1}}\\
    & \Bigg(2 (n+1) \bar p_a (1+m) \bar p_b \sin(\bar a)\sin(\bar b) \Big(2 \beta^2 - (1+\beta^2)(\cos(\bar a) + \cos(\bar b)) \cos(\bar b) \Big) \\
    & - \beta^2 |j_b (1+m) \bar p_b|^{\frac{2}{m+1}} + \frac{1}{4}(1+m)^2 \bar p_b^2 \sin(\bar b)^2  \Big( 4\beta^2 - (1+\beta^2)(\cos(\bar a) + \cos(\bar b))^2 \Big) 
   \Bigg).
\end{aligned}
\end{align}
Since the equations of motion cannot be solved analytically for any values of $n$ and $m$ (within the range of validity established earlier), we will proceed with a numerical analysis of certain choices of values for these parameters and comment on them. 

With the initial conditions specified in the previous section,  the solutions to the equations of motion for the new effective Hamiltonian $H_{\text{eff}}$ can be studied numerically.
As was already pointed out, these initial conditions are only valid for suitably sized black holes for which the Kretschmann scalar at the horizon is well below the critical curvature scale $K_{\text{bh}} \ll K_{\text{crit}}$.
This can be translated into a condition on the black hole mass, namely $M_{\text{bh}}> M_{\text{crit}}$, where the latter is defined from $K_{\text{crit}}$ as given in eq.\,\eqref{eq:Mcrit}.
As per construction, the minimum black hole mass that can be studied with the new model depends on the value chosen for $K_{\text{crit}}$, which is a free parameter of the model and requires further physical input to be determined. 
It is expected, and further shown in the following, that the qualitative behavior of the proposed model is independent of the numerical value of $K_{\text{crit}}$, a particular value only sets a scale of the system.

When establishing key features of the new model, the exponents of the regularization parameters $\Bar{\mu}_a$ and $\Bar{\mu}_b$ (eq.\,\eqref{eq:decoupledMubar}) are set to $n=\frac{1}{2}$ and $m=1$, which correspond to the choice of regularization parameters made in the BMM model  \cite{BMM} and allow for a direct comparison of the results of the two models. 
Other values of $n$ are commented on at the end of this subsection. For the choices of $m$ and $n$ considered, there exists a solution to $H_{\text{eff}} = 0$ after inserting the initial conditions $\bar a^0, \bar p_a^0$ and $\bar p_b^0$ with $|\bar b| \ll 1$
and this is the initial condition chosen for $\bar b^0$.
Thus, for the black holes studied below, the horizon indeed lies in the classical regime where $|\bar a| \ll 1$ and $|\bar b| \ll 1$.
Also, the numerical factors $j_a$ and $j_b$ in the regularization parameters will be set as $j_a = j_b = 1$ and the critical black hole mass is set to one Planck mass  $M_{\text{crit}} = 1\,m_{\text{Pl}}$, which corresponds to setting $K_{\text{crit}} = \frac{3}{4}$ in Planck units.

The lapse in eq.\,\eqref{eq:lapseAnalytic}, while allowing to find an analytical solution for $\bar a(t)$, does not permit to solve the equations of motion numerically as it becomes singular at some point of the evolution and the system cannot be evolved further.
Instead, the lapse chosen to numerically solve the equations of motions is
\begin{equation}
    N = 
     -\frac{\beta}{2} |(n+1) \bar p_a|^{\frac{1}{2(n+1)}} \ |(1+m) \bar p_b|^{\frac{1}{m+1}} \Big((n+1)(1+m) \bar p_b \sin(\bar b)\Big)^{-1}.
     \label{eq:lapseNumerics}
\end{equation}
 
\begin{figure}
    \centering
        \includegraphics[width=0.49\textwidth]{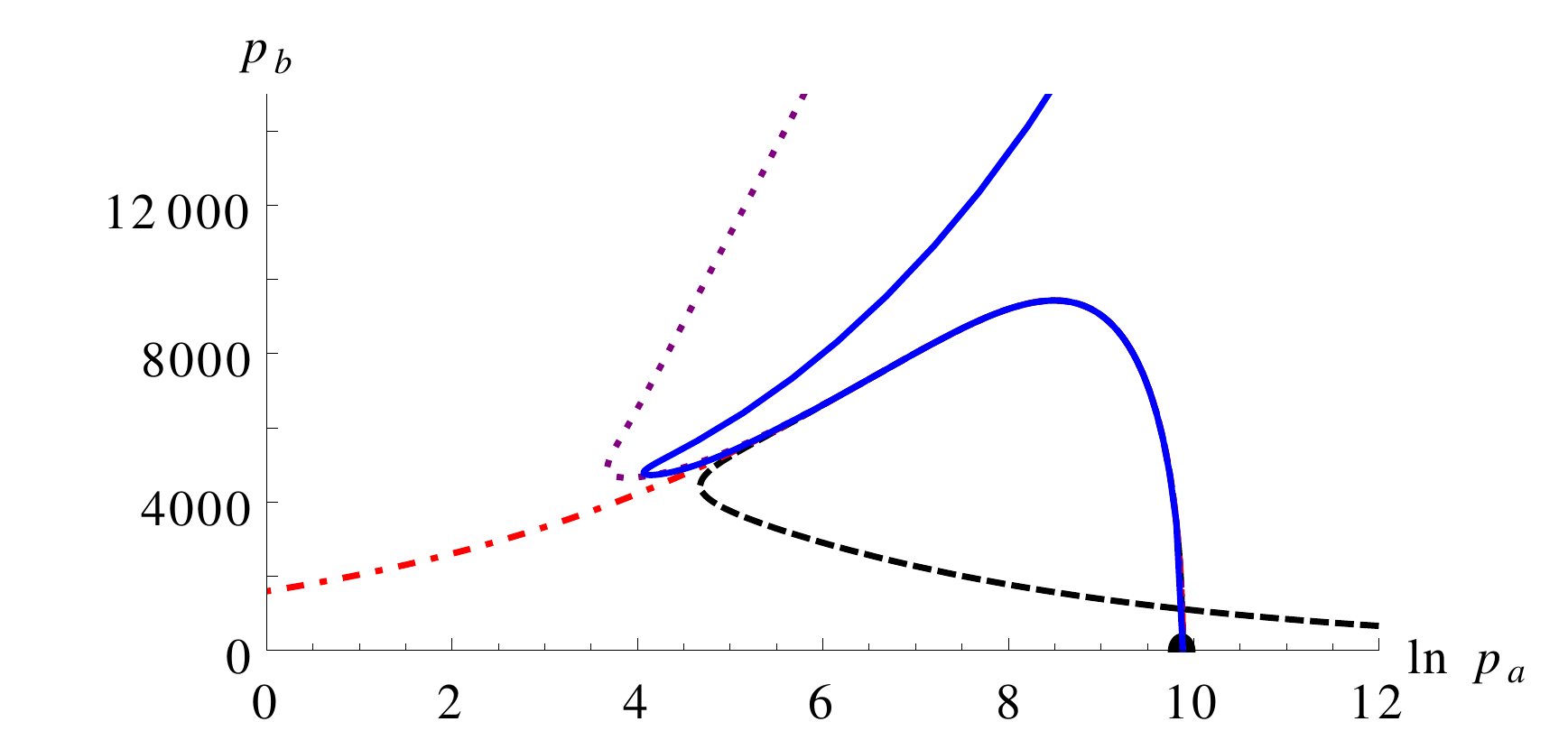}
        \includegraphics[width=0.49\textwidth]{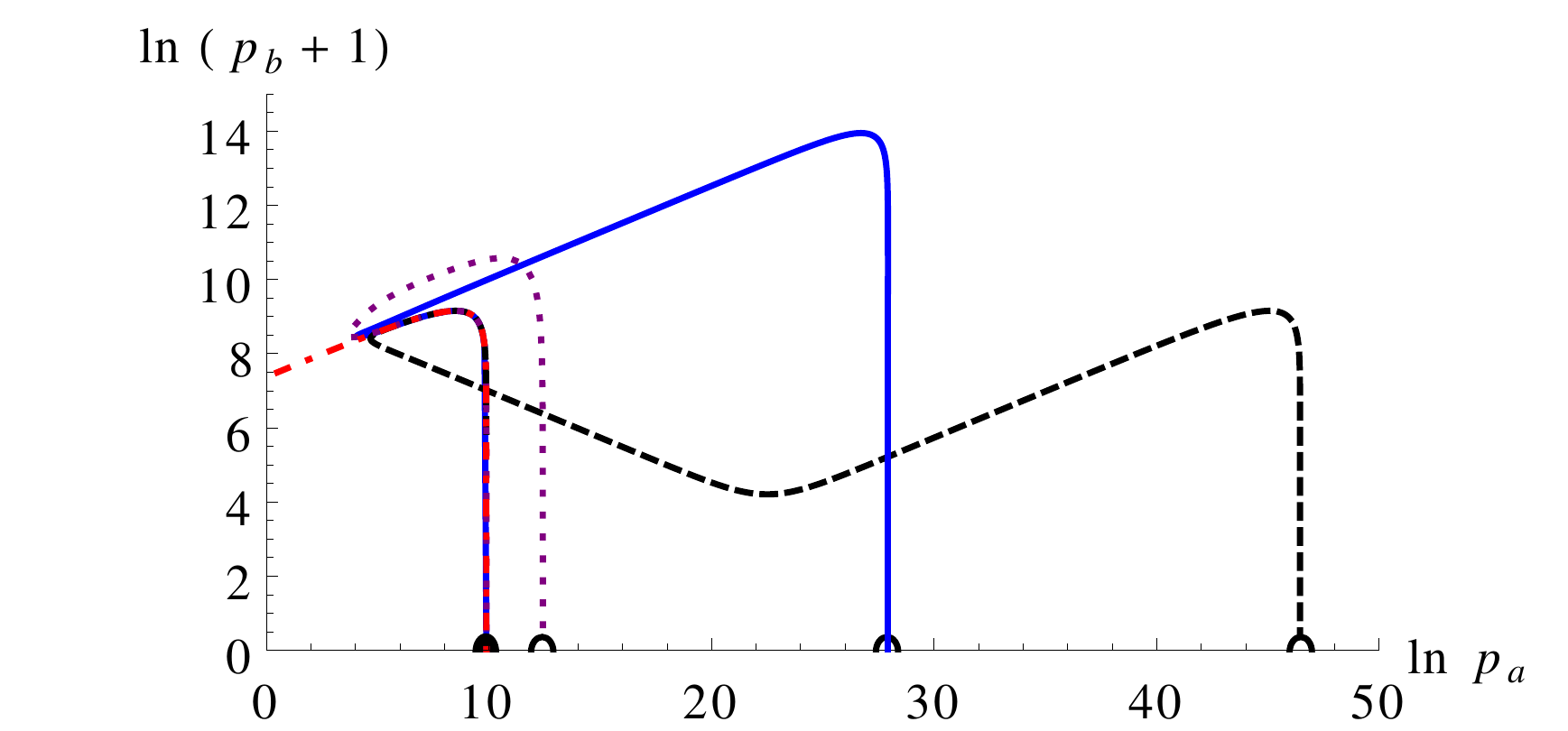}
    \caption{
    Evolution of $p_b$ with respect to $\text{ln}(p_a)$ as generated by $H_{\text{eff}}$ (\textcolor{blue}{blue}, continuous)  with $n=\frac{1}{2}$ and $m=1$, $H_\text{eff}^o$ (\textcolor{purple}{purple}, dotted) with $\mu_{0,a} = \mu_{0,b}= \frac{1}{10}$ \cite{mu0}, the BMM model (black, dashed) \cite{BMM} which corresponds to a ``reduce then regularize'' procedure using the same $\bar \mu$ with $n=\frac{1}{2}$ and $m=1$, and $H_{\text{cl}}$ (\textcolor{red}{red}, dash-dotted).
    These results are obtained for $\beta = 0.2375$ and $ M_{\text{bh}} = 70\, M_{\text{crit}}$.
    Note the logarithmic scale for the vertical axis in the second plot, where the shift by $1$ in the argument was introduced such that the origin of the vertical axis corresponds to $p_b =0$, i.e.\ the two horizons.
    The black hole and white hole horizons are marked by black and white dots respectively.
    }
    \label{fig:compareModelsM70}
\end{figure}

Fig.\,\ref{fig:compareModelsM70} shows the evolution of $p_b$ with respect to $\text{ln}(p_a)$ generated by the new model, compared to general relativity and previously studied models, namely the BMM model  and $H_\text{eff}^o$ (eq.\,\eqref{eq:h2effmu0}). 
This lapse independent depiction allows a comparison to the classical evolution, which is mimicked by the dynamics resulting from $H_{\text{eff}}$ until shortly before $p_a$ reaches its minimum value.
The numerically obtained solutions for $\bar p_a$ and $\bar p_b$ are converted to $p_a$ and $p_b$ in accordance with the definitions in eq.\,\eqref{eq:newvars}.

In the classical Kantowski-Sachs solution, the singularity is encountered at $p_a=0$ where the Kretschmann scalar diverges. Fig.\,\ref{fig:compareModelsM70} shows that a strictly positive minimum value of $p_a$ is present in all depicted effective models, but the exact minimum value depends on the model.
The black hole interior undergoes a transition from a trapped black hole region to an anti-trapped white hole region when $p_a$ reaches its minimum value $p_{a, \rm min}$. 
To establish this, the evolution of the expansion scalar $\theta$ of a congruence of null geodesics is considered. For two future pointing null vectors $k_+^{\mu}$ and $k_-^{\mu}$ corresponding to outgoing and ingoing radial null geodesics respectively, normalized via $k_+^{\mu}k_{\mu \, -} = -1$, the expansion scalar in the Kantowski-Sachs spacetime is given by $\theta_\pm =  \frac{1}{\sqrt{2} |N|}\frac{\dot{p_a}}{p_a}$. Thus the sign of the  expansion scalar depends only on the sign of $\dot{p_a}$. From fig.\,\ref{fig:compareModelsM70}, one can deduce that $\dot{p_a}<0$ inside the black hole horizon prior to $p_{a, \rm min}$, vanishes at the minimum, and subsequently $\dot{p_a}>0$. One can therefore conclude that the minimum of $p_a$ represents a transition surface from a trapped black hole to an anti-trapped white hole region.

After the transition surface, the systems corresponding to the different effective Hamiltonians evolve on largely different scales. The qualitative evolution of the models based on $H_{\text{eff}}$ is the same: after the transition surface, $p_b$ increases up to a second local maximum shortly after which the second horizon occurs. This differs from the BMM model; however, the phenomenologically interesting properties consisting of the singularity resolution, as well as the transition surface and a second horizon, are present in all models. 
The second Killing horizon, marked by $\bar p_b = 0 \Leftrightarrow p_b=0$, appears also in previously studied models of the black hole interior and is dubbed the \emph{white hole horizon}. It occurs at vastly different values of $p_a$ for different models.
The effective model and the solutions to the equations of motion are only valid up to the second horizon. 

\begin{figure}
    \centering
        \includegraphics[width=0.49\textwidth]{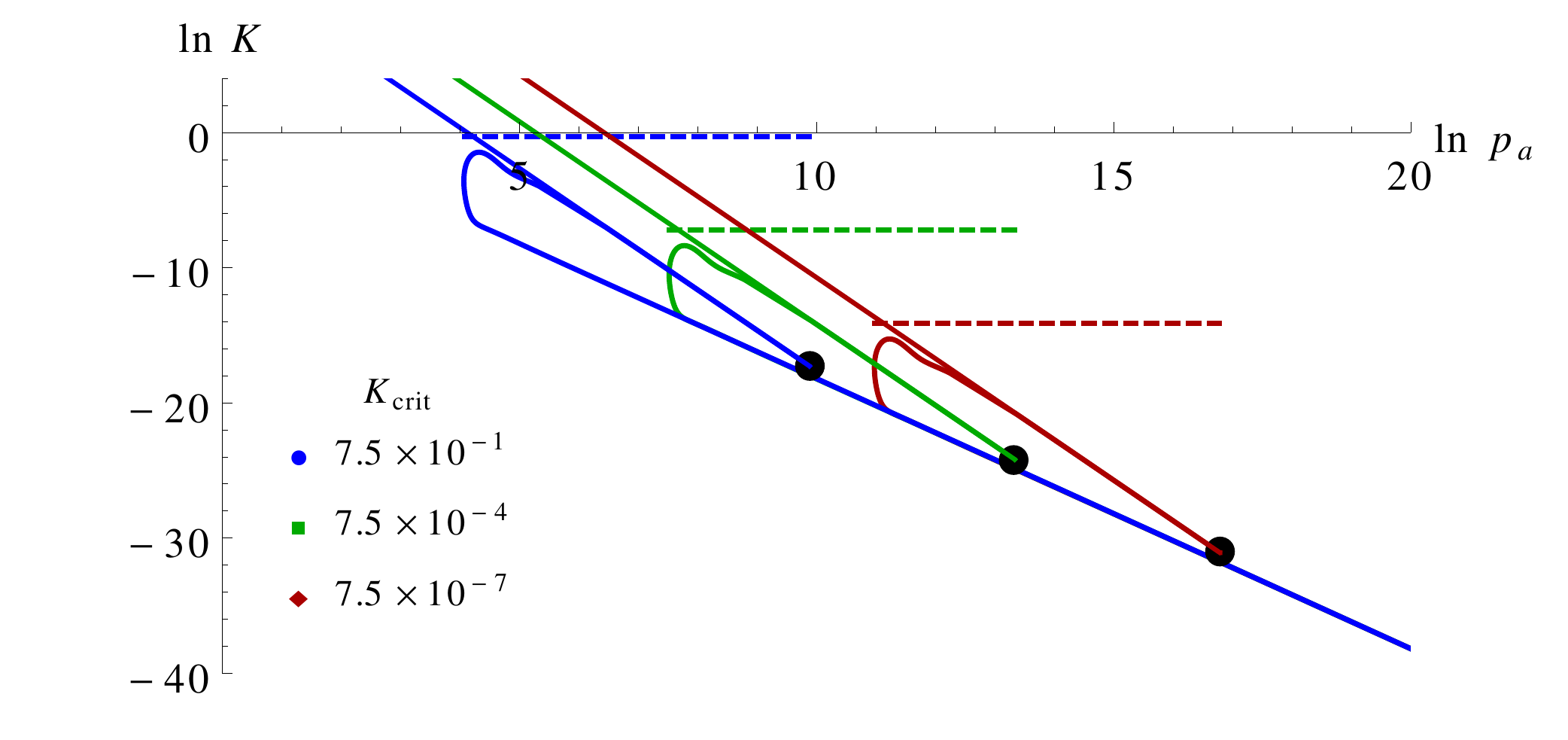}
        \includegraphics[width=0.49\textwidth]{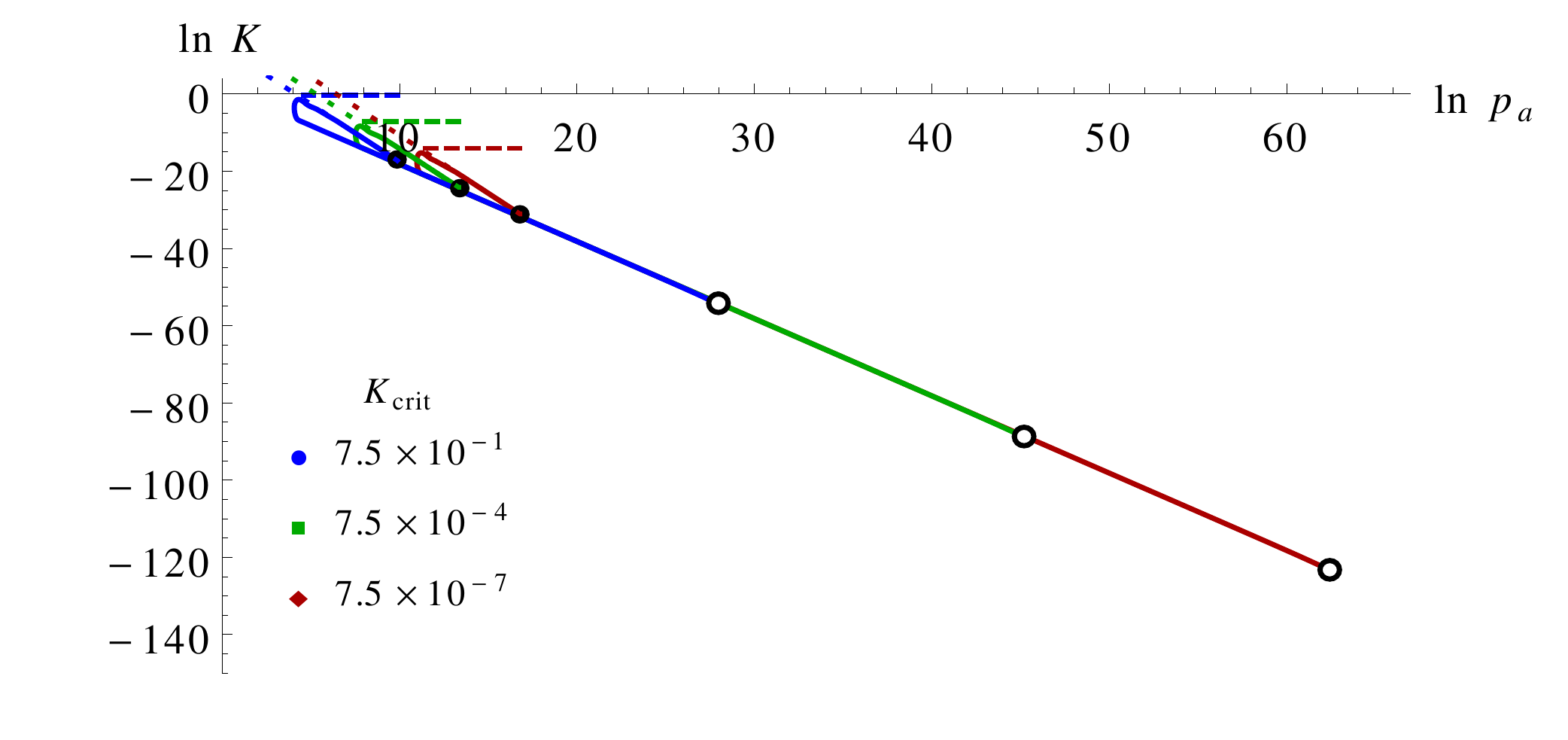}
    \caption{
Logarithmic evolution of the  Kretschmann scalar with respect to $\text{ln}(p_a)$ as generated by $H_{\text{eff}}$ with $n=\frac{1}{2}$ and $m=1$ for different values of $K_{\rm crit}$, where $M_{\rm bh} = 70\,M_{\rm crit}$. 
    The dashed lines indicate the respective value of $K_{\text{crit}}$ and the dotted lines correspond to the evolution of the classical Kretschmann scalar for the respective black hole mass.  
    The black and white dots mark the black hole and white hole horizons respectively.
    We set $\beta = 0.2375$.
    }
    \label{fig:varyKcrit}
\end{figure}

Having established the new model's qualitative properties, it remains to consider the evolution of the Kretschmann scalar in order to verify the presence of a bounce and the consistency of the model dynamics with the considerations made in sec.\,\ref{sec:pheno_init}, when establishing the initial condition $\bar a^0$. 
Fig.\,\ref{fig:varyKcrit} shows the evolution of the Kretschmann scalar with respect to $p_a$ for different values of the critical curvature $K_{\text{crit}}$. The ratio of the black hole mass to the critical mass is kept unchanged with respect to fig.\,\ref{fig:compareModelsM70} $M_{\text{bh}}/M_{\text{crit}}=70$ for the values considered.
There are three important observations: 
\begin{enumerate}
    \item[i.] The effective Kretschmann scalar is bounded from above, confirming that the singularity induced by a classically diverging curvature is lifted (bounce).
    \item[ii.] The effective Kretschmann scalar departs from the classical expression and reaches its maximum value prior to and not at the transition surface ($p_a=p_{a,\text{min}}$). 
    \item[iii.] The evolution is the same, up to a shift, for different values of $K_{\text{crit}}$. This demonstrates that $K_{\text{crit}}$ indeed plays the role of a scale of the system; the behavior of the system is determined by $M_{\text{bh}}/M_{\text{crit}}$.
    This comes as no surprise; $K_{\text{crit}}$ was introduced to play the role of exactly such a scale, and the results from fig.\,\ref{fig:varyKcrit} verify that this is reflected by the model dynamics. 
\end{enumerate}
As was already pointed out in sec.\,\ref{sec:pheno_init}, it is not important that the maximum value of the effective Kretschmann scalar $K_{\text{max}}$ lies above $K_{\text{crit}}$, as long as the black hole horizon can be identified with the classical regime and the region around the bounce with the non-classical regime.
The first is given for sufficiently large black hole masses ($K_{\text{bh}}\ll K_{\text{crit}}$).
The latter is important to ensure that quantum effects do not occur in the classical low curvature regime and is indeed satisfied by the new model: the Kretschmann scalar around the bounce is of a similar order as $K_{\text{crit}}$.
The value of the Kretschmann scalar at the white hole horizon $K_{\text{wh}}$ is very small in comparison to $K_{\text{crit}}$ as can be seen in fig.\,\ref{fig:varyKcrit}. This indicates that the system returns to the classical regime at the white hole horizon, which is of importance in the following, where a notion of a white hole mass is introduced and its dependence on the black hole mass investigated. 

\begin{figure}
    \centering
        \includegraphics[width=0.75\textwidth]{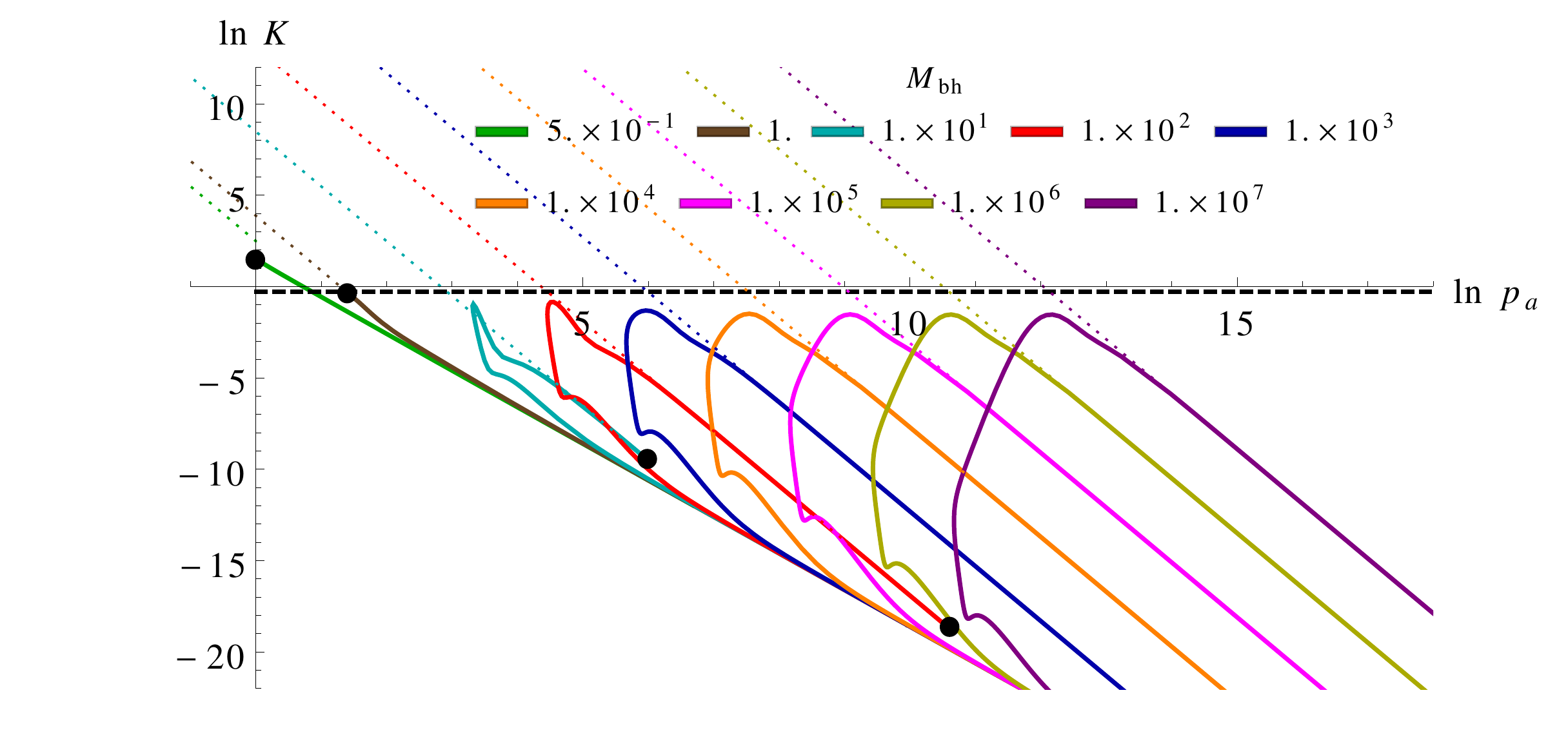}
    \caption{
    Evolution of the  Kretschmann scalar with respect to $\ln (p_a)$ as generated by $H_{\text{eff}}$ with $n=\frac{1}{2}$ and $m=1$ for different black hole masses (given in units of  $M_{\text{crit}} = 1\,m_{\text{Pl}}$, which corresponds to $K_{\text{crit}} = \frac{3}{4}$).
    The dashed black line indicates the value of $K_{\text{crit}}$ and the dotted lines correspond to the evolution of the classical Kretschmann scalar for the respective black hole masses.  
    Where applicable, the black hole horizon is marked with a black dot. In most cases, however, neither the black hole horizon nor the white hole horizon are depicted in the figure; the qualitative evolution of the Kretschmann scalars in the ranges not shown in the plot are the same as in fig.\,\ref{fig:varyKcrit}. 
    We set $\beta = 2.375\times 10^{-3}$.
    }
    \label{fig:KMbh}
\end{figure}

Fig.\,\ref{fig:KMbh} shows the logarithmic evolution of $K(p_a)$ for black hole masses in a range of $0.5 - 10^7\,M_{\text{crit}}$, with $K_{\text{crit}} = \frac{3}{4}$.
For black hole masses of $0.5\,M_{\text{crit}}$ and $1\, M_{\text{crit}}$ the bounce is absent and the effective solutions for $p_a$ and $K$ depart from the classical trajectory immediately inside the horizon.
As the setup of the initial conditions is only valid for sufficiently large black holes, this failure of the model for cases with $K_{\text{bh}}$ of the same order of magnitude as $K_{\text{crit}}$ is expected.
For $M_{\text{bh}} \geqslant 10\,M_{\text{crit}}$, the phenomenological features mentioned previously are recovered and the black hole singularity is resolved. The maximum values of the Kretschmann scalar $K_{\text{max}}$ are of the same order of magnitude as $K_{\text{crit}}$ for the entire mass range, and $K_{\text{max}} \gg K_{\text{bh}}$ holds in all cases. Hence, no quantum effects at low curvature are observed.  Whereas for smaller black hole masses, $K_{\text{max}}$ is reached in close vicinity of the transition surface, the bounce marked by $K_{\text{max}}$ occurs noticeably before $p_{a,\text{min}}$ for larger $M_{\text{bh}}$.
After the bounce, the Kretschmann scalars follow qualitatively the same evolution trajectory until the white hole horizon (but have different final values).
The evolution of $K$ until the white hole horizon is not displayed, but it always mimics the behavior observed in fig.\,\ref{fig:varyKcrit} with $K_{\text{wh}} \ll K_{\text{bh}}$.

Based on the fact that the second horizon is a Killing horizon, one can assign a white hole mass $M_{\text{wh}}$ to this horizon in the same manner as it is done for the black hole, namely $M_{\text{wh}}$ is given by the value of $p_a$ at the white hole horizon\footnote{In absence of a proof of the existence of an asymptotic region beyond the white hole horizon, the quantity in eq.\,\eqref{WHmass} cannot be tied to an asymptotic notion of mass such as the ADM or Bondi masses. However, a direct calculation shows that $M_{\text{wh}}$ in eq.\,\eqref{WHmass} coincides with the value of the Hawking mass associated to the $2$-surface of $t$ and $r$ constant at the second point where $p_b=0$ (i.e. at the white hole horizon). Hence one could use the Hawking mass as a definition of the quantity in eq.\,\eqref{WHmass}, providing a clearer interpretation.} as 
\begin{align}\label{WHmass}
 M_{\text{wh}}:= \frac{\sqrt{p_{a,\text{wh}}}}{2}.
\end{align}
This classical approximation is expected to hold as  $K_{\text{wh}}\ll K_{\text{crit}}$.

\begin{figure}
    \centering
        \includegraphics[width=0.6\textwidth]{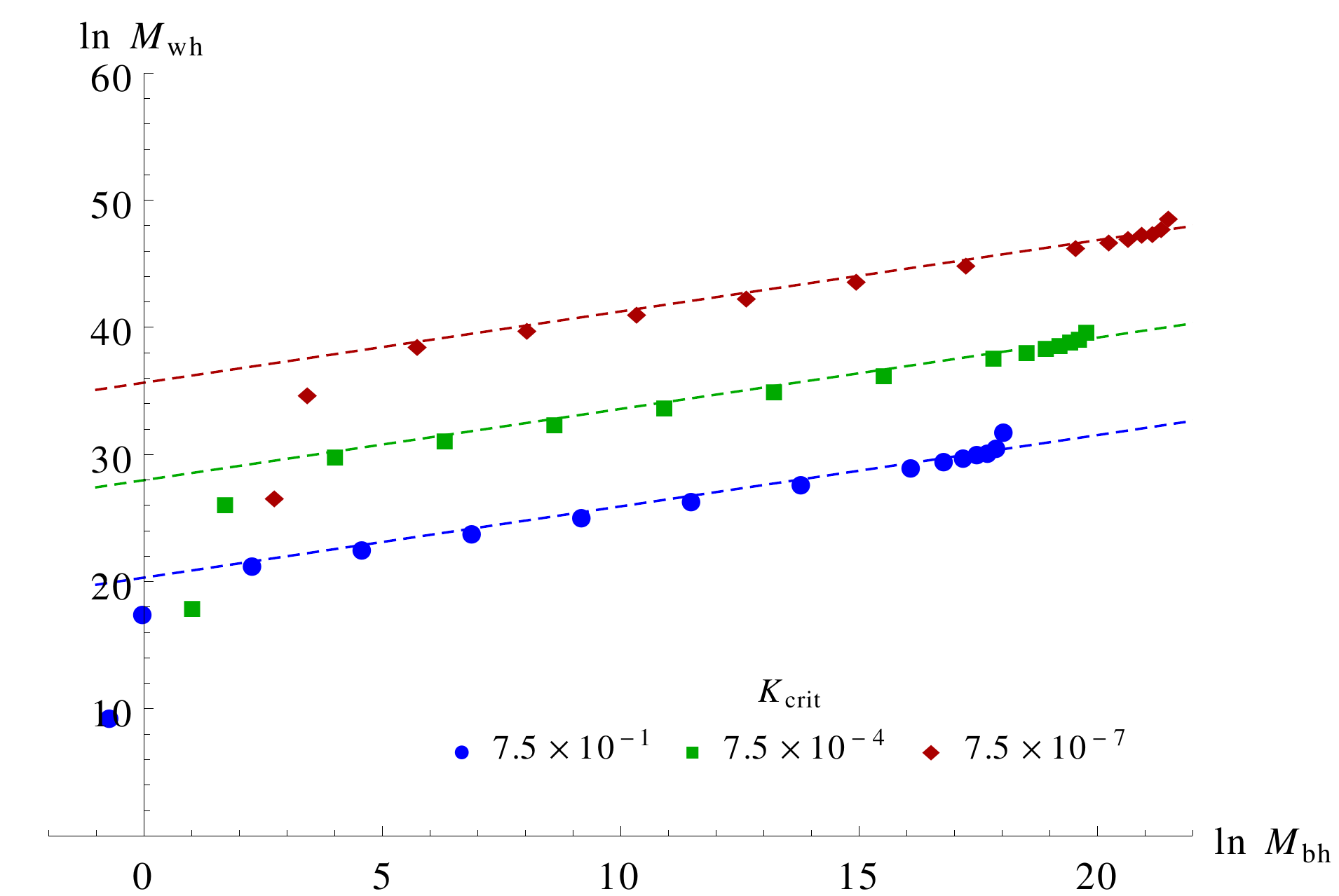}
    \caption{
    The $\text{ln}(M_{\text{wh}})$ as a function of $\text{ln}(M_{\text{bh}})$ for different values of $K_{\text{crit}}$, with $n = \frac{1}{2}$ , $m = 1$ and $\beta = 2.375\times 10^{-3}$. The dashed lines correspond to linear fits $\text{ln}(M_{\text{wh}}) = s \text{ln}(M_{\text{bh}}) + w$ with $s = 0.56$. The fits are carried out for the mass range $M_{\rm bh }= 10\,M_\text{crit} - 10^7\,M_\text{crit}$, after which the relation is clearly non-linear.
    }
    \label{fig:MwhMbhn12}
\end{figure}

Fig.\,\ref{fig:MwhMbhn12} displays the $\text{ln}(M_{\text{wh}})$ as a function of $\text{ln}(M_{\text{bh}})$ for different values of $K_{\text{crit}}$, for the mass range of $M_{\rm bh}$ approximately from $1\,M_\text{crit}$ to $10^8\,M_\text{crit}$. 
The plots confirm that the ratio of the black hole mass to the critical mass $M_{\rm bh}/M_{\rm crit}$ determines the qualitative behavior of the model.
Ignoring the masses outside the validity range of the new model, we observe a linear dependence of $\text{ln}(M_{\text{wh}})$ on $\text{ln}(M_{\text{bh}})$ until $M_{\text{bh}}/ M_{\text{crit}}\approx 10^7$, afterwards the dependence is non-linear. 
A linear fit applied to the mass range $M_{\text{bh}} = 10\,M_{\text{crit}} - 10^7\,M_{\text{crit}}$ gives the relation
\begin{equation}
   \text{ln}(M_{\text{wh}}) \propto 0.56\: \text{ln}(M_{\text{bh}}).
\end{equation}
On one hand, if one extrapolates this relation, one would conclude that for large initial black hole masses, one would observe a generic de-amplification of the black hole mass to obtain a smaller white hole mass. On the other hand, fig.\,\ref{fig:MwhMbhn12} shows a departure from the linear behavior for masses $ M_{\text{bh}}\geqslant10^7$. This non-linearity encountered was studied extensively in an attempt to reveal its numerical accuracy. 
For smaller values of $\beta$, the range of accessible masses and linearity of the mass relation decreases. The fact that $\beta$ influences the non-linear behavior indicates that the breakdown and non-linearity may be of numerical origin. 
However, decreasing the time step during numerical solving and looking for quantifiable errors did not allow to reduce the non-linear behavior, nor define a suitable criterion by which one could rule out the solutions found for large masses. Hence, in absence of analytical solutions, it is not possible at the moment to be unequivocal on the masses relation for an unbounded range of initial black hole masses.

The recovery of the classical regime at the white hole horizon offers the possibility to consider different initial conditions to solve the equations of motion of the system. Namely, one can set the initial conditions at the white hole horizon instead of the black hole horizon. This translates into having
\begin{align}
 \bar p_{a|_\text{wh}} = (4 M_{\text{wh}}^2)^{n+1}/(j_a (n+1)),\qquad \bar a_{|_\text{wh}} = 2\ \text{arccot}\Bigg[\Big(\frac{M_{\text{wh}}}{M_{\text{crit}}}\Big)^{\frac{4}{3}(n+1)}\Bigg].
\end{align}
with $\bar p_{b|_\text{wh}}=0$ and $\bar b_{|_\text{wh}}$ given by the constraint. Since this corresponds to simply replacing $M_\text{bh}$ by $M_\text{wh}$ in the initial conditions, the solutions are the same as the ones obtained earlier, except that now the plots start from the white hole horizon and end at the black hole horizon, in a backward time evolution. In this case, the mass relation is inverted with respect to the one recovered in the case where the initial conditions are set at the black hole horizon, and one then gets an amplification of the black hole mass. This should not be surprising as this could be inferred from the time reversal symmetry satisfied by the equations of motion, meaning that given a solution to the equations of motion, its time reversal image is also a solution.

The reader shall be reminded at this point that the new model permits any  choice of regularization parameters as given in eq.\,\eqref{eq:decoupledMubar}, with the restrictions $m\leqslant1$ and $n>-1$. The model parameters $n=\frac{1}{2}$ and $m=1$ used so far were chosen to allow a comparison of the results with the BMM model. The case $m=1$ furthermore simplifies the equations of motion; its influence on the model is commented briefly at the end of this section. As $n$ explicitly appears in the classicality condition eq.\,\eqref{eq:classCondK}, and therefore in the expression for $\bar a^0$, it is interesting to look at different values of $n$. 

\begin{figure}
    \centering
        \includegraphics[width=0.7\textwidth]{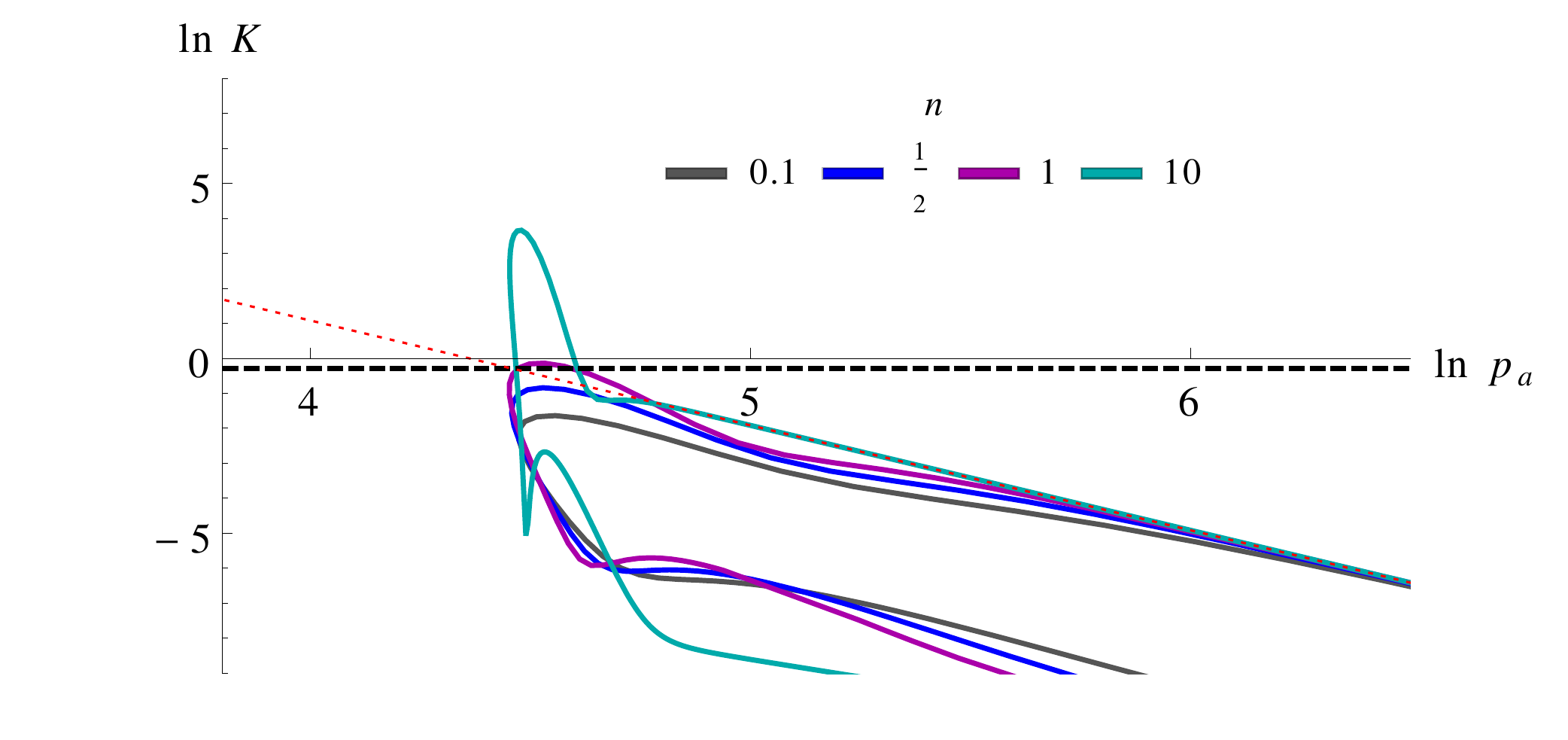}
    \caption{
    Logarithmic evolution of the Kretschmann scalar with respect to $\text{ln}(p_a)$ as generated by $H_{\text{eff}}$ for different values of $n$, with $ m=1$, $K_{\text{crit}} = \frac{3}{4}$ and $M_{\text{bh}}= 100\,M_{\text{crit}}$. 
    The dashed black line indicates the value of $K_{\text{crit}}$ and the dotted line correspond to the evolution of the classical Kretschmann scalar for the same given black hole mass. For clarity, the plots are restricted to a range which does not include the black hole and white hole horizons; beyond this range, the Kretschmann scalars qualitatively follow an evolution as depicted in fig.\,\ref{fig:varyKcrit}.     
    We set $\beta = 2.375\times 10^{-3}$.
    }
    \label{fig:Knall}
\end{figure}

When taking various values of $n$, one actually observes that the characteristic properties of the model are not different from the $n=\frac{1}{2}$ case, namely one recovers a bounce, transition surface, and a second horizon. Fig.\,\ref{fig:Knall} shows the evolution of the effective Kretschmann scalar for different values of $n$ with $M_{\text{bh}} = 100\, M_{\text{crit}}$ and $K_{\text{crit}} = \frac{3}{4}$.
For each value of $n$, the upper bound on $K$ is approximately mass independent, similarly to what we have seen in fig.\,\ref{fig:KMbh}, but $K_{\text{max}}$ increases with increasing $n$. 

\begin{figure}
    \centering
    \includegraphics[width=0.6\textwidth]{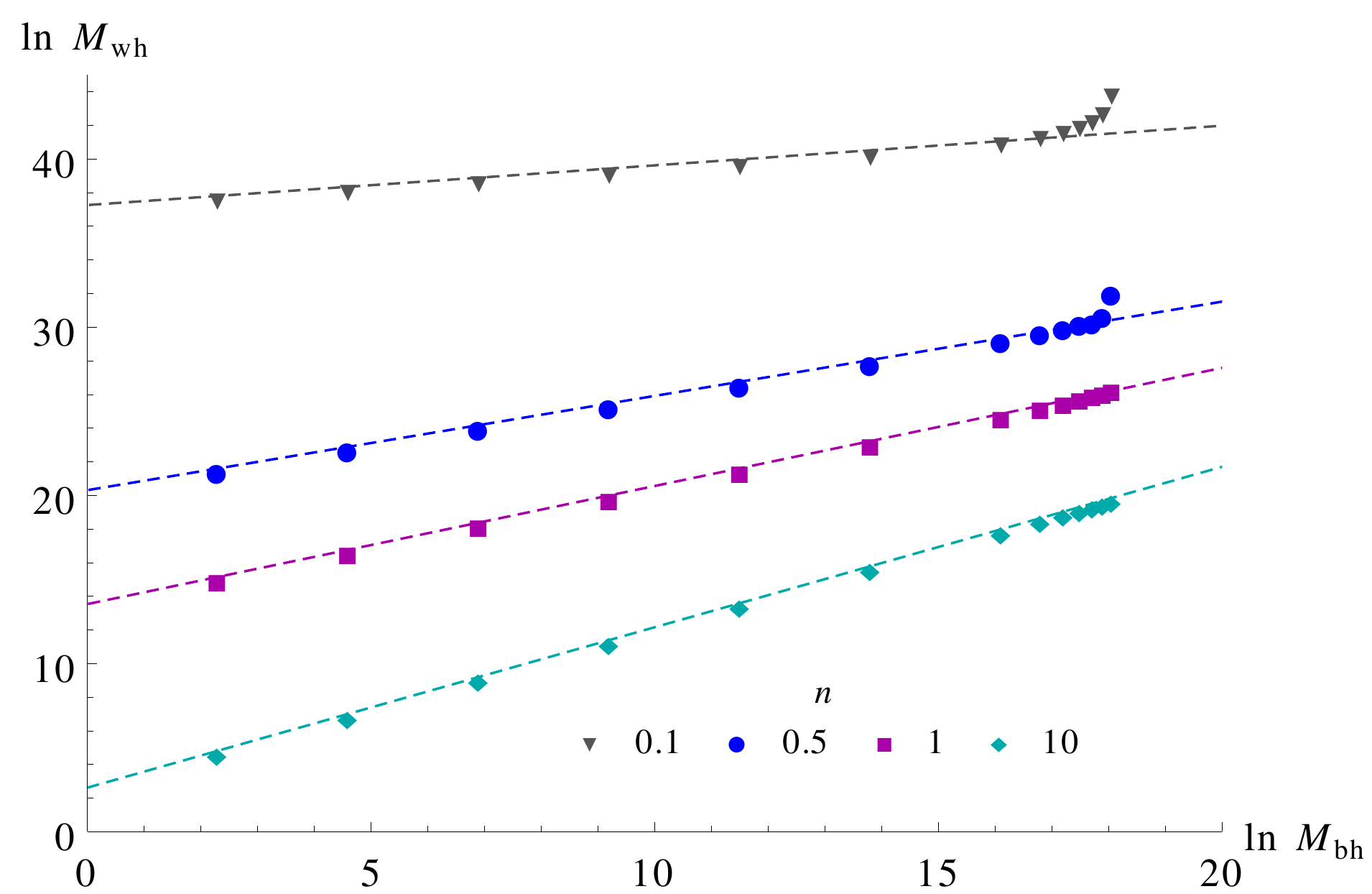}
     \caption{
    Relation between the white hole and black hole mass for different values of $n$ and $m=1,\ M_{\text{crit}} = 1\,m_{\text{Pl}}$ on a logarithmic scale.
    The dashed lines correspond to linear fits of the logarithmic dependence of the white hole mass on the black hole mass,  $\text{ln}M_{\text{wh}} = s\, \text{ln}M_{\text{bh}} + w$, with
     $s \cong 0.24;\: 0.56;\: 0.70;\: 0.95$ for $n= 0.1;\: \frac{1}{2};\: 1;\: 10$ respectively. The fits were carried out for the mass range $M_{\text{bh}} =  10- 10^7\,M_{\text{crit}}$. We set $K_{\text{crit}} = \frac{3}{4}$ and $\beta = 2.375\times 10^{-3}$.
    }
    \label{fig:MwhMbhnall}
\end{figure}

The relations between the black hole and white hole masses are depicted in fig.\,\ref{fig:MwhMbhnall}, where the linear fit for  $\text{ln}(M_{\text{wh}})\big[\text{ln}(M_{\text{bh}})\big]$ is again carried out for $M_{\text{bh}} = 10\,M_{\text{crit}} - 10^7\,M_{\text{crit}}$.
The slopes $s$ of the fits, $ \text{ln}(M_{\text{wh}}) \propto s\, \text{ln}(M_{\text{bh}})$, are given by approximately $s \approx 0.24;\: 0.70;\: 0.95$ for $n= 0.1;\: 1;\: 10$ respectively.
This shows that the mass relation depends on $n$; it shall also be noted that the non-linear behavior for masses above $10^7\,M_{\text{crit}}$ decreases with increasing $n$. 

Previously studied models of the black hole interior found different relations between the white hole and black hole masses. In the AOS model, the symmetric case ($s=1$) was obtained, while in the BMM model mass relations with $s = \frac{3}{5}$ and $s=\frac{5}{3}$ were introduced based on considerations for physically viable models.
A dependence of the mass relation on the choice of $a^0$, which is classically an irrelevant quantity, was already observed in previous models \cite{mu0, BMM}, and is encountered also in the model we study through the choice of $n$.
In the new model, a linear relation of the form $\text{ln}(M_{\text{wh}}) \propto s\, \text{ln}(M_{\text{bh}})$ is recovered for a limited range of the black hole mass only. 
For the values of $n$ probed, a mass de-amplification $s<1$ is observed, whereas in the case of $n=10$ the slope of the fit is close to one ($s\cong 0.95$). Larger values of $n$ lead to extremely large and small initial values $\bar p_a^0$ and $\bar a^0$ respectively, which causes complications in the numerical solving. Further investigation is needed to study mass relations for large $n$ and establish whether the symmetric case $s=1$ can be achieved in the new model.  
A non-linear mass relation in the high mass range as is observed for larger $n$ has not been encountered in any of the other models; but it could so far not be resolved or attributed to numerical effects to a satisfactory level.

To conclude this section, we comment on the variation of the exponent $m$ in the regularization parameter $\bar{\mu}_b$, while $n=\frac{1}{2}$ is fixed.
The choice $m=1$ as used in previous sections simplifies the equations of motion, and when determining $\bar b^0$ from the constraint $H_{\text{eff}}=0$ using the initial conditions $\bar p_b^0,\ \bar p_a^0,\ \bar a^0$ and  the lapse eq.\,\eqref{eq:lapseNumerics}, there is a solution for which $\bar b$ is close to but not exactly zero.
For $m<1$, $\bar b=0$ is an exact solution (which is no longer apparent after inserting the lapse). It is the only solution for which $|\bar b|\ll 1$ at the horizon, hence $\bar b^0=0$ for $m<1$.
This however leads to divergences in the equations of motion at the black hole horizon, 
and other choices of lapse cause further complications in the numerical solving. 
Knowing that the effective model matches the classical evolution also just inside the horizon, one can circumvent this issue by setting the initial conditions from the classical expressions (eq.\,\eqref{class-analytical-sol}) close to $t_0$.
The validity of this approximation is confirmed by establishing that the numerical results  still agree with the classical solutions for the first part of the evolution.
For the values of $m$ probed ($m=\frac{1}{2}$, $m=\frac{1}{4}$, and $n=\frac{1}{2}$), a bounce and second horizon are recovered. Unlike in the case of $m=1$, the solutions for $\bar p_a$  and $\bar p_b$ are discontinuous at the white hole horizon and the white hole mass cannot be calculated from the value of $\bar p_a$ at the horizon directly.
Instead, it is approximated from  $\bar p_a$ shortly before the second horizon is reached.
Considerations on permissible values of $m$ as done in sec.\,\ref{sec:newM_mu}  do not rule out negative values. 
First numerical tests indicate that $m<0$ might lead to a change in model phenomenology, but further investigation is needed.

\section{Conclusions and outlook}

In this article, we developed a new loop effective model for the Schwarzschild black hole interior using a regularization procedure based on Thiemann identities and a family of phase space dependent regularization parameters $\Bar{\mu}_a$ and $\Bar{\mu}_b$.
When constructing the new model, a certain class of phase space dependent regularization parameters was identified, for which the regularized versions of Thiemann identities remain valid in order to recover the correct classical limit.

The complexity of the equations of motion did not allow for the derivation of analytical solutions, hence we proceeded with a numerical analysis in which we chose the Kretschmann scalar $K$ to play the role of a scale in the evolution of the system. Consequently, a universal critical curvature scale $K_{\text{crit}}$ was introduced, to which we associate a critical initial black hole mass $M_{\text{crit}}$, arising as an additional parameter in the model which must be fixed based on physical considerations. Those critical parameters set a range of validity for our effective model, namely the initial black hole mass satisfies $M_{\text{bh}} \geqslant 10\,M_{\rm crit}$. 

The analysis shows that for the initial black hole masses probed within the range of validity, we observe a resolution of the classical black hole singularity, induced by the existence of an upper bound on the Kretschmann scalar $K_{\text{max}}$, while keeping the curvature at the horizon $K_{\text{bh}}$ in the classical regime. Given a critical curvature $K_{\text{crit}}$ many orders of magnitude larger than $K_{\text{bh}}$, the value of $K_{\text{max}}$ is at a scale comparable to $K_{\text{crit}}$, satisfying the criterion that quantum effects become relevant at large curvatures. Past this point where $K_{\text{max}}$ is reached, which we call the bounce, the system evolves up to a second Killing horizon. 
Post bounce, there is a transition surface from the trapped black hole region to an anti-trapped region, the ``white hole''-like region. Thanks to the fact that the Kretschmann curvature at the white hole horizon lies in the classical regime, we were able to associate a mass to this second horizon and we investigated the relation between the white hole mass and black hole mass and its dependence on the regularization parameters. 
Across several orders of magnitude of the black hole mass, a linear mass relation of the form $\ln(M_{\text{wh}}) = s\, \text{ln}(M_{\text{bh}}) + w$ was established. 
In fact, using the classicality condition at both horizons, one identifies two sets of initial conditions characterizing one set of solutions and its image via time reversal. The two sets differ at the level of the mass relation, where one set corresponds to a black hole mass de-amplification $s<1$, while the image set corresponds to a black hole mass amplification $s>1$. However, it remains to be shown whether the symmetric case $s=1$ can be recovered for the family of regularization parameters we studied.

An aspect in which the new model differs significantly from previous findings is that the mass relation seems to no longer follow the aforementioned linear dependence for very large black holes. 
The non-linearity is accompanied by a breakdown of the numerical solving, i.e.\ there exists a black hole mass beyond which the numerical solving fails, and the value of this mass depends only on the value of the critical mass $M_{\rm crit}$. The origin of this issue could not be determined in a satisfactory manner, and may require a different approach to obtain a conclusive result.

Finally, it would be desirable to develop a quantum model of the black hole interior with the new Hamiltonian, and establish whether the effective model studied in this article could be derived from the quantum theory. This may shed more light on the mass relation and perhaps elucidate the problem arising in the numerical analysis in the effective model. We hope to tackle these questions in a future work.

\section*{Acknowledgements}
This work was supported by the German Research Foundation (DFG) under the project BA 4966/1-2 and Germany's Excellence Strategy EXC 2121 “Quantum Universe” - 390833306, as well as by the Polish National Science Center OPUS 15 Grant nr 2018/29/B/ST2/01250.

\bibliographystyle{ieeetr}
\bibliography{bib}
\end{document}